\documentclass[prd,12pt,nofootinbib,superscriptaddress]{revtex4-2}

\usepackage{bm}
\usepackage{amsmath}
\usepackage{amssymb}
\usepackage{graphicx}
\usepackage{subfigure}
\usepackage{hyperref}
\usepackage{color}
\usepackage{comment}
\usepackage{ulem}
\usepackage{CJK}
\usepackage{orcidlink}
\usepackage{array}
\usepackage{indentfirst}
\usepackage{float}

\hypersetup{
	colorlinks=true,
	linkcolor=red,
	citecolor=blue,
}
\newcommand{\Xin}{X^{\textup{in}}_{l m\omega}} 
\newcommand{\Xup}{X^{\textup{up}}_{l m \omega}} 

\begin{document}

\begin{CJK*}{UTF8}{gbsn}
\title{Probing new fundamental fields with Extreme Mass Ratio Inspirals} 	

\author{Chao Zhang  \orcidlink{0000-0001-8829-1591}}
\email{zhangchao666@sjtu.edu.cn}
\affiliation{Institute of Fundamental Physics and Quantum Technology, Department of Physics, School of Physical Science and Technology, Ningbo University, Ningbo, Zhejiang 315211, China}

\author{Yungui Gong  \orcidlink{0000-0001-5065-2259}}
\email{Corresponding author. gongyungui@nbu.edu.cn}
\affiliation{Institute of Fundamental Physics and Quantum Technology, Department of Physics, School of Physical Science and Technology, Ningbo University, Ningbo, Zhejiang 315211, China}

\begin{abstract}
We examine extreme mass ratio inspirals (EMRIs), where a charged compact object spirals into a supermassive black hole, in modified gravity theories with additional scalar or vector fields.
Using the Teukolsky and generalized Sasaki-Nakamura formalisms,
we provide the post-Newtonian expansion of the energy flux of the vector waves up to $O(v^5)$ beyond the quadrupole formula in the weak field and numerically calculate the energy flux in the strong field for a charged particle moving in circular orbits.
Our findings reveal a degeneracy in the scalar and vector charge parameters for weak-field, slow-motion orbits.
However, for strong-field, fast-motion orbits close to the innermost stable circular orbit, we observe distinct behaviors between scalar and vector fields. 
We investigate the potential of using EMRIs detected by space-based gravitational-wave detectors, such as the Laser Interferometer Space Antenna to identify whether a black hole carries a scalar or vector charge.
The influence of scalar and vector flux on the orbital evolution and tensor GW phase is not suitable for us to distinguish scalar and vector fields for their correlations that exist between the scalar and vector flux.
However, extra polarizations emitted by the scalar or vector field can break the correlations between the scalar field and vector field and then help us distinguish the scalar and vector field.
\end{abstract}

\maketitle
\end{CJK*}

\section{Introduction}

The accelerated expansion of the late Universe and the dark matter phenomenon in cosmology indicate the existence of extra fields such as scalars \cite{ Berti:2015itd, Essig:2013lka, Hui:2016ltb, Barack:2018yly, Barausse:2020rsu, Bolton:2022hpt} and vectors \cite{ Arias:2012az, Hambye:2008bq, DiChiara:2015bua,Frieman:2008sn, Yoo:2012ug, Copeland:2006wr}.
Introducing such fields are also the most natural generalization of general relativity, regarded as a practical field theory with a cutoff at the Planck scale \cite{Arkani-Hamed:2002bjr}.
Although the strong constraints on these new fundamental fields and modified gravity have been provided by the Solar system and binary pulsar tests in the weak gravitational fields, the strong-field regime tests are still needed \cite{Will:1993hxu, Will:2005va}.
Fortunately, upcoming space-based gravitational-wave detectors such as the Laser Interferometer Space Antenna (LISA) \cite{Danzmann:1997hm,Audley:2017drz}, TianQin \cite{TianQin:2015yph} and Taiji \cite{Hu:2017mde, Gong:2021gvw} offer us with potential to detect scalar fields \cite{Barsanti:2022vvl, Guo:2022euk, Maselli:2021men, Maselli:2020zgv, Barsanti:2022ana, Yunes:2011aa, Cardoso:2011xi, Zhang:2022rfr,AbhishekChowdhuri:2022ora} and vector fields \cite{Torres:2020fye, Zhang:2022hbt, Zhang:2023vok, Liang:2022gdk,Bhattacharyya:2023kbh} and deviation from black hole \cite{Ghosh:2024arw,AbhishekChowdhuri:2023rfv,Rahman:2023sof,Rahman:2021eay,Rahman:2022fay,Rahman:2023sof,AbhishekChowdhuri:2023rfv,AbhishekChowdhuri:2023gvu,Kumar:2024utz}. 
However, it remains to be seen whether these two types of fields can be distinguished experimentally.
The additional energy loss channels from extra fields cause a noticeable accumulated imprint on standard polarizations.
This effect can be considered compact objects carrying a "charge" \cite{Maselli:2020zgv}.
The difference in energy loss caused by scalar fields and vector fields determines the distinction between them for detectors.
In the previous research literature \cite{Cardoso:2020iji, Liu:2020cds, Liu:2020vsy}, the scalar and vector fields appeared indistinguishable because their emitted fluxes share the same behavior falling of $v^8$ caused by the dipole radiation at low velocity $v\ll 1$.
The vector flux in the case of slow motion has the form
\begin{equation}
 \dot{E}_{q_{\rm vector}}=\frac{2q_{\rm vector}^2m_p^2}{3M^2}v^8,
\end{equation}
while the scalar flux behaves like
\begin{equation}
\dot{E}_{q_{\rm scalar}}=\frac{q_{\rm scalar}^2m_p^2}{3M^2}v^8,
\end{equation}
where $q_{\rm scalar}$ and $q_{\rm vector}$ are the scalar and vector charges, respectively.
If we redefine the charge with a constant factor such as $q_{\rm vector}=q_v$ and $q_{\rm scalar}=\sqrt{2}q_s$, then the scalar and vector charges cannot be distinguished for the same emission contribution to the GW phase.
However, it is still unknown whether this degeneration persists beyond the leading, Newtonian-dipole formula in the weak field at higher post-Newtonian (PN) expansion or in the strong field region.
Therefore, our goal is to investigate whether scalar and vector fields can be distinguished during the long inspiral of binary systems that can be detectable by space-based GW detectors.

The general action, including massless or sufficiently light scalar field as is the case in scalar-tensor theories \cite{Campbell:1991kz, Mignemi:1992nt,Kanti:1995vq,Yunes:2011we,Kleihaus:2011tg,Sotiriou:2013qea,Sotiriou:2014pfa,Antoniou:2017acq,Doneva:2017bvd,Silva:2017uqg,Cardoso:2020iji}, massless or sufficiently vector charge as is the case in some dark matter models with mutli-charged components \cite{Holdom:1985ag,Cardoso:2020iji,Cardoso:2016olt} has the form
\begin{equation} 
S[g_{\mu\nu},\Phi,A_\mu,\Psi]=S_0[g_{\mu\nu},\Phi,A_\mu]+\alpha S_c[g_{\mu\nu},\Phi]+\beta S_v[g_{\mu\nu},A_\mu]+S_{\rm m}[g_{\mu\nu},\Phi,A_\mu,\Psi]\,,\label{action}
\end{equation}
where
\begin{equation}
  \label{S_0}
S_0= \int d^4x \frac{\sqrt{-g}}{16\pi}\left(R-\frac{1}{2}\partial_\mu\Phi\partial^\mu\Phi-\frac{1}{4}F^{\mu\nu}F_{\mu\nu}\right)\,,
\end{equation}
$g$ is the metric determinant, $R$ is the Ricci scalar, $\Phi$ is a massless scalar field, $A_\mu$ is a massless vector field, $F_{\mu\nu}=\nabla_{\mu}A_{\nu}-\nabla_{\nu}A_{\mu}$ is the field strength and $\Psi$ is the matter field.
$\alpha S_c$ describes the non-minimal coupling between the metric and scalar fields, while $\beta S_v$ represents the non-minimal coupling between the metric and vector fields.
Binary systems can effectively detect new fields for high curvature during an inspiral-merger-ringdown period.
However, calculation of GWs based on numerical relativity is computationally expansive for binary systems in the strong field region.
Among them, extreme mass ratio inspirals (EMRIs) offer a significant simplification in modeling the signal in the presence of new fundamental fields and in providing insight into how deviations in the signal arise \cite{Maselli:2021men}.
EMRIs consist of a stellar-mass compact object (secondary object) with mass $m_p\sim1-100~M_{\odot}$ such as BHs, neutron stars, white dwarfs, etc. orbiting around a supermassive black hole (SMBH) (primary object) with mass $M\sim10^5-10^7~M_{\odot}$.
As the secondary body orbits the SMBH, its motion will deviate from the geodesic curve by emitting fluxes.
Apart from the emitted flux for the standard GW polarizations, scalar or vector radiation is emitted throughout the inspiral due to the accelerating secondary body with charges emitting radiation such as dipole emission.
Theoretically, the deviations of BH in general relativity (GR) and the interactions between gravity and new fields are more conspicuous at larger curvatures, and the charge of BH is also significant.
Several effects, such as superradiance \cite{Brito:2015oca, Arvanitaki:2009fg, Cardoso:2011xi}, spontaneous scalarization \cite{Damour:1993hw, Silva:2017uqg, Doneva:2017bvd}, or spontaneous vectorization \cite{Barton:2021wfj, Oliveira:2020dru, Ramazanoglu:2017xbl}, induce BHs or compact stars with different structures than their GR counterparts, making new fields detectable only in the surroundings of compact objects.
For very different characteristic curvatures near the horizon as one over the mass squared in the EMRI system, the primary object can be taken as chargeless while only the secondary object carries charges \cite{Maselli:2020zgv,Zhang:2022hbt}.
Furthermore, the standard GW emission is decoupled from extra fields, and its flux for the polarizations of the emitted GWs will be virtually unaffected even if it carries a scalar charge or vector charge \cite{Maselli:2020zgv,Zhang:2023vok}.

In this paper, we consider the EMRI system that a stellar-mass-charged BH inspires into a Kerr SMBH.
We obtain the evolution of the orbit by implementing the BH perturbation method and then construct the waveforms.
The ability to distinguish scalar and vector fields for detectors is calculated by the dephasing, faithfulness, the Fisher information matrix method, and Bayesian analysis.
We take LISA as a representative of the space-based detectors in our discussion,
and the analysis can be easily extended to other space-based detectors.
The paper is organized as follows. 
In Sec.~\ref{method}, we introduce the basic formalism of the BH perturbation method and calculate the energy flux carried by the scalar and vector field.
Then, we show the analytical post-Newtionian fluxes and compare them with the numerical results in Sec.~\ref{flux}.
In Sec.~\ref{DF}, we calculate the dephasing and faithfulness caused by the scalar and vector field.
The results of parameter estimation with the Fisher matrix and Bayesian methods are shown in Sec.~\ref{FIMB}. 
In Sec.~\ref{EPT}, we add the extra polarizations emitted by the scalar and vector field to the signals.
Last, we summarize in Sec.~\ref{conclusion}.

\section{Method}\label{method}
Due to its charge, the secondary body acts as a scalar or vector field equation source.
In the setup described in the introduction, the perturbed scalar field equation takes the simple form \cite{Maselli:2020zgv}
\begin{equation}
     \square\Phi=4\pi \sqrt{2} q_s m_p\int \frac{\delta^{(4)}(x-z(\tau))}{\sqrt{-g}}d\tau,
\end{equation}
Following the same procedures \cite{Maselli:2020zgv}, the perturbed vector field equation takes the form
\begin{equation}
\nabla_\nu F^{\mu\nu}=4\pi q_v m_p\int ~ u^{\mu}\frac{\delta^{(4)}\left[x-z(\tau)\right]}{\sqrt{-g}}d\tau,
\end{equation}
where $u^{\mu}$ is the velocity of the secondary object and $\tau$ is the proper time.
The functions $z^\mu(\tau)$ describe the geodesic curve of a point particle moving in the exterior Kerr spacetime.
The orbital energy of the particle decreases due to the total energy emission,
\begin{equation}
\dot{E}_{\rm orb}=-\dot{E}_\text{grav}-\dot{E}_{\rm vec}-\dot{E}_{\rm scal},
\end{equation}
where $\dot{E}_{\text{grav}}$, $\dot{E}_{\rm vec}$ and $\dot{E}_{\rm scal}$ represents the gravitational flux, vector flux, and scalar flux, respectively.
So the scalar and vector flux affects the orbital evolution of the secondary body.
By computing the emitted energy flux, we can determine the adiabatic evolution of the EMRI.
Since the orbital evolution determines the GW phase, the scalar or vector field emission contributes to the GW phase of the EMRI.

Extending a method developed by Sasaki in the Schwarzschild case and by Shibata, Sasaki, Tagoshi, and Tanaka in the Kerr case \cite{Shibata:1994jx, Tagoshi:1996gh, Hughes:2000pf, Ohashi:1996uz, Sasaki:2003xr}, we calculate the PN expansion of the vector flux in a circular orbit around a rotating black hole up to $O(v^5)$.
In the Newman-Penrose formalism, a single master equation for tensor ($s=-2$), vector ($s=-1$), and scalar ($s=0$) perturbations was derived as \cite{Teukolsky:1973ha},
\begin{equation}
\label{TB}
\begin{split}
&\left[\frac{(r^2+a^2)^2}{\Delta}-a^2\sin^2{\theta}\right]\frac{\partial^2\psi}{\partial t^2}+\frac{4ar}{\Delta}\frac{\partial^2\psi}{\partial t\partial\varphi}\\
&\quad+\left[\frac{a^2}{\Delta}-\frac{1}{\sin^2{\theta}}\right]\frac{\partial^2\psi}{\partial \varphi^2}-\Delta^{-s}\frac{\partial}{\partial r}\left(\Delta^{s+1}\frac{\partial\psi}{\partial r}\right)\\
&\qquad-\frac{1}{\sin\theta}\frac{\partial}{\partial \theta}\left(\sin\theta\frac{\partial\psi}{\partial \theta}\right)-2s\left[\frac{a(r-1)}{\Delta}+\frac{i\cos\theta}{\sin^2{\theta}}\right]\frac{\partial\psi}{\partial \varphi}\\
&\qquad\qquad-2s\left[\frac{(r^2-a^2)}{\Delta}-r-ia\cos\theta\right]\frac{\partial\psi}{\partial t}\\
&\qquad\qquad\qquad\qquad\qquad+(s^2\cot^2\theta-s)\psi=4\pi\varSigma T,
\end{split}
\end{equation}
where $a$ is the dimensionless angular momentum, $\Delta=r^2-2r+a^2$ and $\varSigma=r^2+a^2\cos^2{\theta}$.
The general formalism of the source term $T$ for different spin $s$ can be seen in the literature \cite{Teukolsky:1973ha}.
The details of the source term for spin $s=0$ on the equatorial circular orbit can be seen in literature \cite{Maselli:2021men,Maselli:2020zgv,Guo:2022euk,Barsanti:2022ana}.
The details of the source term for spin $s=-1$ on the equatorial circular orbit can be seen in literature \cite{Zhang:2022hbt,Zhang:2023vok,Torres:2020fye,Torres:2020fye}.
The details of the source term for spin $s=-2$ on the equatorial circular orbit can be seen in literature \cite{Cutler:1994pb,Sasaki:2003xr,Sago:2015rpa}.
In terms of the eigenfunctions ${_{s}}S_{lm}(\theta)$ \cite{Teukolsky:1973ha,Goldberg:1966uu}, the field $\psi$ can be written as   %
\begin{equation}
\psi=\int d\omega \sum_{l,m}R_{\omega lm}(r)~{_{s}}S_{lm}(\theta)e^{-i\omega t+im\varphi},
\end{equation}
where the radial function $R_{\omega lm}(r)$ satisfies the inhomogeneous Teukolsky equation
\begin{equation}
\label{Teukolsky}
\Delta^{-s}\frac{d}{d r}\left(\Delta^{s+1}\frac{d R_{\omega lm}}{d r}\right)-V_{T}(r)R_{\omega lm}=T_{\omega lm},
\end{equation}
the function
\begin{equation}
V_{T}=-\frac{K^2-2is(r-1)K}{\Delta}-4is\omega r+\lambda_{lm\omega},
\end{equation}
 $K=(r^2+a^2)\omega-am$, $\lambda_{lm\omega}$ is the corresponding eigenvalue which can be computed by the BH Perturbation Toolkit \cite{BHPToolkit}.
The solution of radial function is purely outgoing at infinity $(+)$ and purely ingoing at the horizon $(-)$,
\begin{equation}
\begin{split}
R_{\omega lm}(r\to r_+)=\delta(\omega-m \hat{\omega})\Delta^{-s}\mathcal{A}^{(+)}_{\omega lm} e^{-i\kappa r^*},\\
R_{\omega lm}(r\to \infty)=\delta(\omega-m \hat{\omega})r^{-2s-1}\mathcal{A}^{(-)}_{\omega lm}e^{i\omega r^*},
\end{split}
\end{equation}
where $\kappa=\omega-m a/(2r_+)$, $r_\pm=1\pm\sqrt{1-a^2}$, $\hat{\omega}$ is the orbital angular frequency, and the tortoise radius of the Kerr metric
\begin{equation}
r^*=r+\frac{2r_+}{r_+-r_-}\ln \frac{r-r_+}{2}-\frac{2r_-}{r_+-r_-}\ln \frac{r-r_-}{2}.
\end{equation}
The full solution allows computing the total energy flux radiated by the EMRIs:
\begin{equation}
    {\dot E}_{\rm orb}=-\sum_{i=+,-}{\dot E}^{(i)}_{\rm grav}+{\dot E}^{(i)}_{\rm scal}+{\dot E}^{(i)}_{\rm vec}\ ,
\end{equation}
where the gravitational flux at infinity  and the horizon can be computed as
\begin{equation}
\dot{E}_{\text{grav}}^{(i)}=\sum_{l=2}^{\infty}\sum_{m=1}^{l}\alpha^{(i)}_{lm}\frac{|\mathcal{A}^{(i)}_{\omega lm}|^2}{2\pi\omega^2},
\end{equation}
A similar formula for vector and scalar fields reads
\begin{equation}
    \begin{split}
{\dot E}^{(i)}_{\rm vec}&=\sum_{l=1}^{\infty}\sum_{m=1}^{l}\beta^{(i)}_{lm}\frac{|\mathcal{A}^{(i)}_{\omega lm}|^2}{\pi},\\
{\dot E}^{(i)}_{\rm scal}&=\sum_{l=1}^{\infty}\sum_{m=1}^{l}\gamma^{(i)}_{lm}|\mathcal{A}^{(i)}_{\omega lm}|^2,
    \end{split}
\end{equation}
the coefficients $(\alpha^{(i)},\beta^{(i)},\gamma^{(i)})$ are given in \cite{Teukolsky:1974yv}.
In this section, we provide further technical details on the formalisms that we compute the PN expansion of the vector flux.
The solutions of the homogeneous Teukolsky and generalized Sasaki-Nakamura (GSN) equations \cite{Hughes:2000pf} are related by
\begin{equation}
\label{eq:fromSNtoTeu}
\begin{split}
R_{lm\omega}(r)&=\frac{\left(
\alpha + \beta_{,r} \Delta^{s+1}\right) \chi -\beta\Delta^{s+1} \chi'}{\eta},   \\
\chi&=\frac{X}{{\sqrt{(r^2 + a^2)\Delta^s}}},
\end{split}
\end{equation}
for $s=-1$,
\begin{equation}
\label{eq:em_alpha_beta}
\begin{split}
\alpha &= \frac{(r^2 + a^2)}{r^2}\sqrt{\Delta}\left[ - {\frac{r}{r^2 + a^2}} -
\frac{i K}{\Delta}\right],\\
\beta &= \frac{(r^2 + a^2)}{r^2}\sqrt{\Delta},
\end{split}
\end{equation}
and $X$ satisfies the GSN equation
\begin{equation}
\frac{d^2X}{dr^{*2}} - {_{s}}F(r) \frac{dX}{dr^*} - {_{s}}U(r) X = 0.
\label{eq:SNeq}
\end{equation}
The coefficient ${_{s}}F(r)$ is
\begin{equation}
{_{s}}F(r)= \frac{\eta(r)_{,r}}{\eta(r)}\frac{\Delta}{r^2 + a^2},
\end{equation}
the function ${_{s}}U(r)$ is
\begin{equation}\label{eq:gsn_potentials}
\begin{split}
{_{s}}U(r)=& {_{s}}U_1(r)\frac{\Delta}{(r^2 + a^2)^2}+ {_{s}}G(r)^2 \\
&+
\frac{d{_{s}G(r)}}{dr}\frac{\Delta}{r^2 + a^2} -
\frac{\Delta{_{s}G(r)}{_{s}F_1(r)}}{r^2 + a^2},
\end{split}
\end{equation}
\begin{eqnarray}
{_{s}}F_1(r) &=& \frac{\eta(r)_{,r}}{\eta(r)},
\end{eqnarray}
\begin{equation}
\begin{split}
{_{s}}U_1(r) =& V_T + \frac{1}{\beta\Delta^s}
\left[\left(2\alpha + \beta_{,r}\Delta^{s+1}\right)_{,r} \right.\\
&\qquad\qquad\qquad\left.- \frac{\eta(r)_{,r}}{\eta(r)}
\left(\alpha + \beta_{,r}\Delta^{s+1}\right)\right],
\end{split}
\label{eq:F1_and_U1}
\end{equation}
\begin{equation}
{_{s}}G(r) = \frac{r\Delta}{(r^2 + a^2)^2} + \frac{s(r - M)}{r^2 + a^2},
\label{eq:G_function}
\end{equation}
where ${}_{,r}$ denotes the derivative with respect to $r$ and
\begin{equation}
\eta(r) = c_0 + \frac{c_1}{r} + \frac{c_2}{r^2} + \frac{c_3}{r^3} + \frac{c_4}{r^4},
\end{equation}
where
\begin{equation}
\label{eq:em_eta_cofs}
\begin{split}
c_0 =& -\lambda_{lm\omega},\\
c_1 =& -2iam,\\
c_2 =& a^2(1 - 2\lambda_{lm\omega}),\\
c_3 =& -2a^2(1 + i a m),\\
c_4 =& a^4(1 - \lambda_{lm\omega}).
\end{split}
\end{equation}
The generalized Sasaki-Nakamura equation admits two linearly independent solutions, $\Xin$ and $\Xup$, with the asymptotic behaviour
\begin{equation}
\Xin
\sim
\begin{cases}
e^{-i \kappa r^\ast} \quad  &r \to r_+  \\
A^{\textup{out}}_{l m \omega}  e^{i \omega r^\ast} + A^{\textup{in}}_{l m
\omega} e^{-i \omega r^\ast} \quad
&\hat{r} \to \infty  \label{eq:inBCSN}
\end{cases}\,,
\end{equation}
\begin{equation}
\Xup\sim
\begin{cases}
 C^{\textup{out}}_{l m \omega} e^{i \kappa r^{\ast}} + C^{\textup{in}}_{l m
\omega}e^{- i \kappa r^{\ast}} \quad \,  &r\to r_+   \label{eq:upBCSN} \\
e^{i \omega r^{\ast}}  \, \quad &r \to \infty
\end{cases}\,.
\end{equation}

\section{Fluxes}\label{flux}
Now we consider the PN expansion of $\Xin$ characterized by the ingoing-wave boundary condition at the horizon to calculate the luminosity for the vector field at infinity.
First we introduce the variable $z=\omega r\sim v$, $z_\pm=\omega r_\pm$ and $\epsilon=2\omega\sim v^3$, then convert $\Xin$ into $\xi_{lm}$ as
\begin{equation}
\label{xi}
\Xin=\sqrt{z^2+a^2\omega^2}\xi_{lm}(z)e^{-i\chi(z)},
\end{equation}
where
\begin{equation}
    \chi(z)=\omega r^*-z-\frac{\epsilon}{2}ma\frac{1}{z_+-z_-}\ln \frac{z-z_+}{z-z_-}.
\end{equation}
Inserting Eq. \eqref{xi} and the angular eigenvalue $\lambda_{lm\omega}=\lambda_{lm\omega}^{(0)}+a\omega\lambda_{lm\omega}^{(1)}+a^2\omega^2\lambda_{lm\omega}^{(2)}+O(v^9)$ into Eq. \eqref{eq:SNeq} and expanding it in terms of $\epsilon$, we obtain
\begin{equation}
    L^{(0)}[\xi_{\ell m}]=\epsilon L^{(1)}[\xi_{\ell m}]
+\epsilon Q^{(1)}[\xi_{\ell m}]+\epsilon^2 Q^{(2)}[\xi_{\ell m}]
+O(\epsilon^3),
\end{equation}
where differential
operators $L^{(0)}$, $L^{(1)}$, $Q^{(1)}$ are given by
\begin{equation}
    \begin{split}
     L^{(0)}&={d^2\over dz^2}+{2\over z}{d\over dz}+\left(1-{\lambda_{lm\omega}^{(0)}
\over z^2}\right), \\
 L^{(1)}&={1\over z}{d^2\over dz^2}
+\left( {{1}\over z^2}
+{2i\over z}\right){d\over dz}
-\left( {1\over z^3}-{i\over z^2}+{1\over z}  \right), \\
Q^{(1)}&={{ia\lambda_{lm\omega}^{(1)}}\over 2z^2}{d\over dz}
-{{ima}\over {\lambda_{lm\omega}^{(0)}z^3}}
+{ma\over {\lambda_{lm\omega}^{(0)}z^2}}+{{\lambda_{lm\omega}^{(1)}a+2ma}\over 2z^2},
    \end{split}
\end{equation}
and we don't list $Q^{(2)}$ here for its long expression.
Expanding $\xi_{\ell m}$ in terms of $\epsilon$ as
\begin{equation}
    \xi_{\ell m}=\sum_{n=0}^{\infty}\epsilon^n \xi^{(n)}_{\ell m}(z),
\end{equation}\label{eps}
we obtain $\xi_{lm}$  from  the iterative equations,
\begin{equation}
\begin{split}
   &L^{(0)}[\xi^{(0)}_{\ell m}]=0, \\
   &L^{(0)}[\xi^{(1)}_{\ell m}]=L^{(1)}[\xi^{(0)}_{\ell m}]
+Q^{(1)}[\xi^{(0)}_{\ell m}]\equiv W^{(1)}_{\ell m},
\\
&L^{(0)}[\xi^{(2)}_{\ell m}]=L^{(1)}[\xi^{(1)}_{\ell m}]
+Q^{(1)}[\xi^{(1)}_{\ell m}]+Q^{(2)}[\xi^{(1)}_{\ell m}]
\equiv W^{(2)}_{\ell m}.
\end{split}
\end{equation}
After complicated calculations, we have
\begin{equation*}
\begin{split}
X^{\textup{in}}_{1 m\omega} =&\frac{z^2}{3}-\frac{z^4}{30}+\frac{z^6}{840}+\epsilon  \left(\frac{i a m}{6}  z-\frac{6+ia m}{40} z^3 +\frac{89+9 i a m}{7560}z^5\right)\\
&+\epsilon ^2 \left(\frac{3 a^2 m^2+3 a^2-2 i a m}{24}+ \frac{4i a^2 m^2-ia^2-4  a m}{24} z \right),
\end{split}
\end{equation*}
\begin{equation*}
\begin{split}
X^{\textup{in}}_{2 m\omega} =&\frac{z^3}{15}-\frac{z^5}{210}+\epsilon  \left(\frac{-27216+4536 i a m}{544320}z^2\right.\\
&\left.+\frac{-7992-396 i a m}{544320} z^4 +\frac{546+13 i a m}{544320}z^6 \right),
\end{split}
\end{equation*}
\begin{equation*}
\begin{split}
X^{\textup{in}}_{3 m\omega} &=\frac{z^4}{105}.
\end{split}
\end{equation*}
Once we have the homogeneous solutions of the GSN equation, we can easily perform the transformation \eqref{eq:fromSNtoTeu} to obtain the corresponding solutions of the Teukolsky equation \eqref{Teukolsky}.
Following the method developed by Sasaki in the Schwarzschild case and by Shibata, Sasaki, Tagoshi, and Tanaka in the Kerr case,
The luminosity for the vector field at infinity is given by
\begin{equation}\label{vpn}
\begin{split}
\dot{E}^\infty_{\text{vec}}=\frac{2q_v^2}{3}&\left(\frac{m_p}{M}\right)^2v^8\left\{1-2 v^2+ (2\pi -4 a) v^3 \right.\\
&\left.+\left(a^2-\frac{61}{10}\right) v^4+\left(\frac{2 \pi }{5}-\frac{103 a}{12}\right) v^5\right\},
\end{split}
\end{equation}
while the scalar field luminosity at the infinity is \cite{Ohashi:1996uz}
\begin{equation}\label{spn}
\begin{split}
\dot{E}^\infty_{\text{scal}}=\frac{2q_s^2}{3}\left(\frac{m_p}{M}\right)^2v^8&\left\{1-2 v^2+ (2\pi -4 a) v^3\right.\\
&\left.+\left(a^2-10\right) v^4+\left(\frac{12 \pi }{5}+4a\right) v^5\right\}.
\end{split}
\end{equation}
The PN expansion up to $O(v^3)$ yields identical scalar field and vector field fluxes.
This degeneracy for the scalar field and vector field in the weak field prevents us from determining the type of charge carried by BHs.
Differences in the behaviors of scalar and vector fields emerge at higher PN expansions and may be distinguishable for fast-motion orbits close to the innermost stable circular orbit (ISCO) in the strong-field region.
However, the PN expansion of luminosities fails in strong gravitational regions, and we need numerically calculate the energy fluxes for both gravitational, scalar, and vector perturbations induced by a charged particle moving in equatorial circular orbits by using the Teukolsky and generalized Sasaki-Nakamura formalisms for the gravitational, scalar, and vector perturbations about a Kerr BH.

\begin{figure}
    \centering
    \includegraphics[width=0.85\columnwidth] {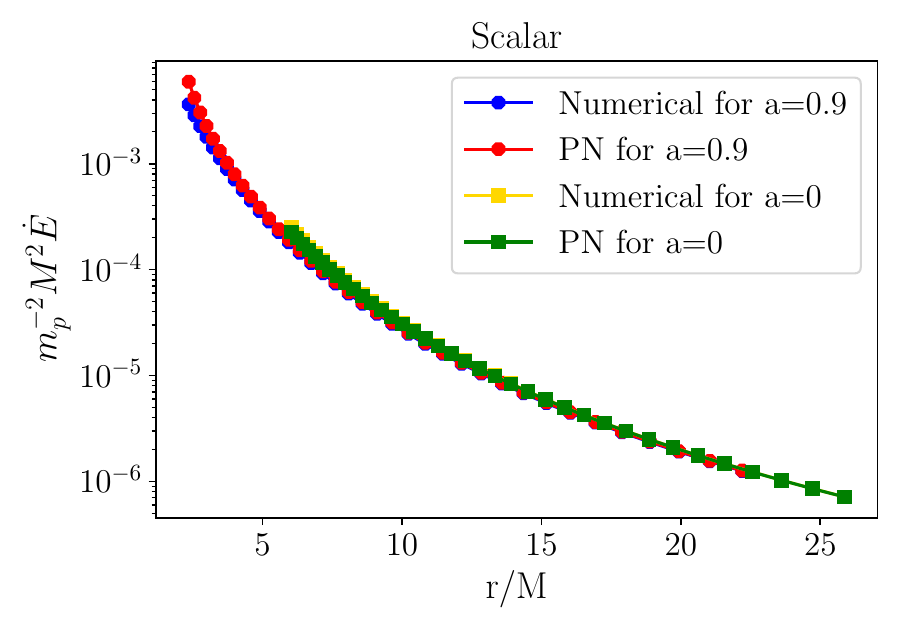}\\
     \includegraphics[width=0.85\columnwidth] {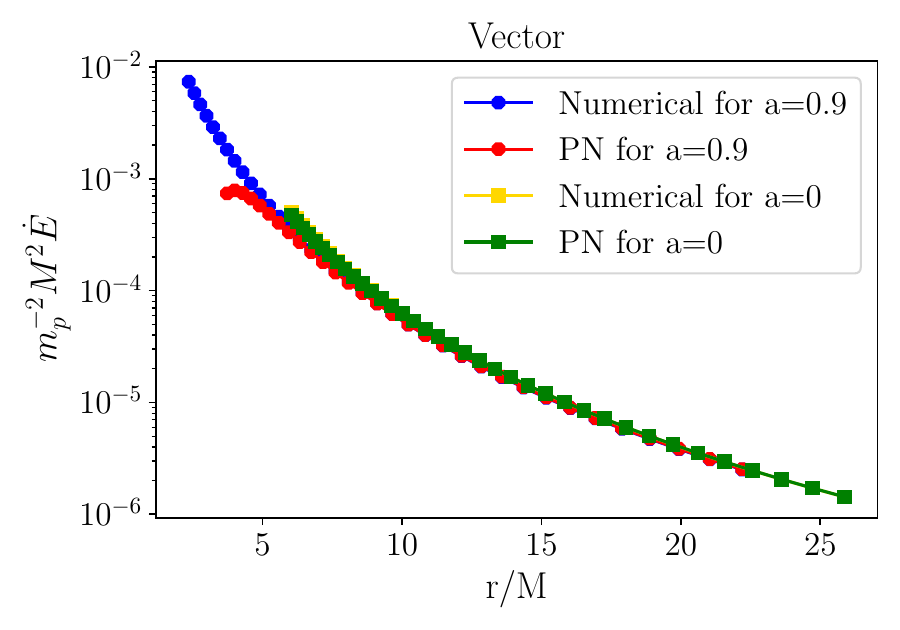}
    \caption{
    The scalar and vector radiation flux from a charged particle on a circular orbit around a BH of spin $a$ as a function of the radius $r$.
    The term "PN" represents the result of PN expansion, and the term "Numerical" represents the fluxes calculated by the Teukolsky formalisms for perturbations of a BH.}
    \label{energy}
\end{figure}

\begin{figure}
    \centering
    \includegraphics[width=0.96\columnwidth]{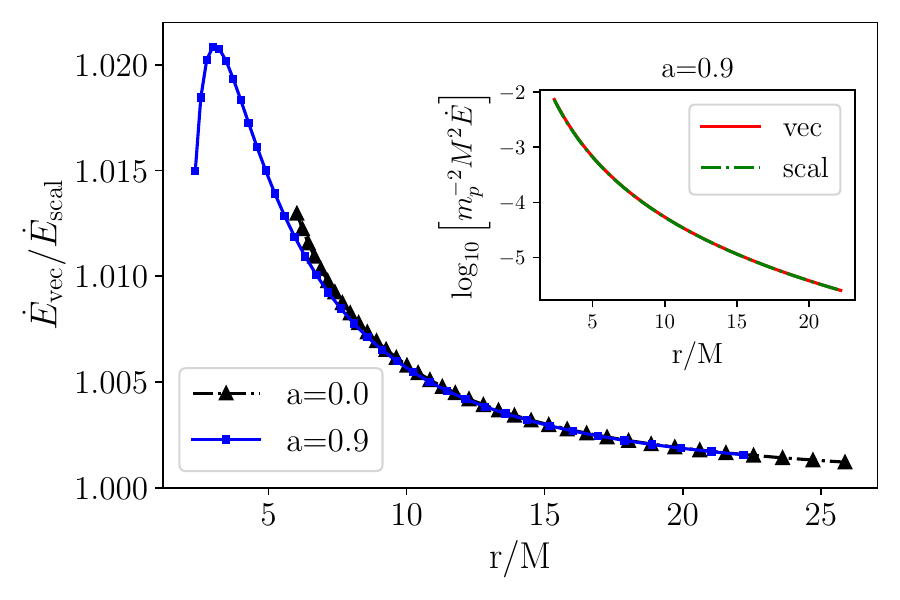}
    \caption{
    The ratio of radiation flux between a vector field and scalar field from a charged particle on a circular orbit around a BH of spin $a$ as a function of the radius $r$. The black dashed line represents the ratio for the Schwarzchild black hole, and the solid blue line indicates the Kerr BH with $a=0.9$.
    The inset shows the normalized flux values for the vector and scalar fields in the unit charge.
    }
    \label{energyratio}
\end{figure}

We compute the energy flux of the gravitational field, vector field, and scalar field, summing all the multipole contributions up to $l_{\rm max}=15$.
We have compared the numerical results of our code against known results published in the literature.
The relative difference of energy flux for different EMRI configurations can be seen in Table \ref{comparision}.
\begin{table}
  \centering
  	\begin{tabular}{|c|c|c|c|c|}
		\hline
$(a,r)$ & \multicolumn{2}{|c|}{(0.5,\ 8.0)} & \multicolumn{2}{|c|}{(0.0,\ 8.0)} \\ \hline
& This paper& Reference   & This paper& Reference\\ \hline
$M^2\dot{E}_{\text{scal}}^{\infty}/m^2_p$ & $6.894(4)\times10^{-5}$& $6.894(5)\times10^{-5}$ \cite{Warburton:2010eq}& $7.637(2)\times10^{-5}$& $7.637(4)\times10^{-5}$ \cite{Warburton:2010eq}\\ \hline
$M^2\dot{E}_{\text{scal}}^{H}/m^2_p$ & $-1.324(5)\times10^{-6}$& $-1.325(3)\times10^{-6}$ \cite{Warburton:2010eq}& $8.823(0)\times10^{-7}$& $8.807(0)\times10^{-7}$    \cite{Warburton:2010eq}            \\ \hline
$M^2\dot{E}_{\text{vec}}^{\infty}/m^2_p$ & $1.388(0)\times10^{-4}$& / &  $1.541(0)\times10^{-4}$& / \\ \hline
$M^2\dot{E}_{\text{vec}}^{H}/m^2_p$ &$-2.487(5)\times10^{-6}$& / & $1.671(0)\times10^{-6}$& /                \\ \hline
	\end{tabular}
    \caption{Comparison of the results on the energy flux.
    For a given $(a,r)$, we calculate the energy $\dot{E}_{\text{scal}}$ carried away by the scalar field with the scalar charge $q_{\rm scalar}=\sqrt{2}q_s=1$, the energy $\dot{E}_{\text{vec}}$ carried away by the vector field with the vector charge $q_{\rm vector}=1$, respectively.
    }
    \label{comparision}
\end{table}
Figure \ref{energy} shows the normalized scalar and vector flux $m_p^{-2}M^2\dot{E}$ as a function of radius for different spins $a=0.9$ and $a=0.0$.
From Fig.~\ref{energy}, the behavior of the scalar and vector fluxes in the slow motion case with the velocity $v\ll 1$ is consistent with the PN results in Eqs.~\eqref{vpn} and \eqref{spn}.
Figure \ref{energyratio} shows the ratio of the energy flux for the vector field and scalar field as a function of the radius, while the inset provides the value of $\dot E_{\rm scal}$ and $\dot E_{\rm vec}$ for Kerr BH with dimensionless $a=0.9$.
For a smaller radius (higher velocities) in the strong field, the ratio of vector flux to scalar flux exceeds $1$ by at most $2\%$, indicating the potential for distinguishing the vector and scalar field.
For larger spin $a$, the ISCO is smaller, and the difference in flux between the scalar field and vector field is more significant.
It implies that the Kerr BH background will more efficiently distinguish the scalar and vector field by space-based detectors.

\section{Dephasing and Faithfulness}\label{DF}

\begin{figure}
    \centering
    \includegraphics[width=0.96\columnwidth]{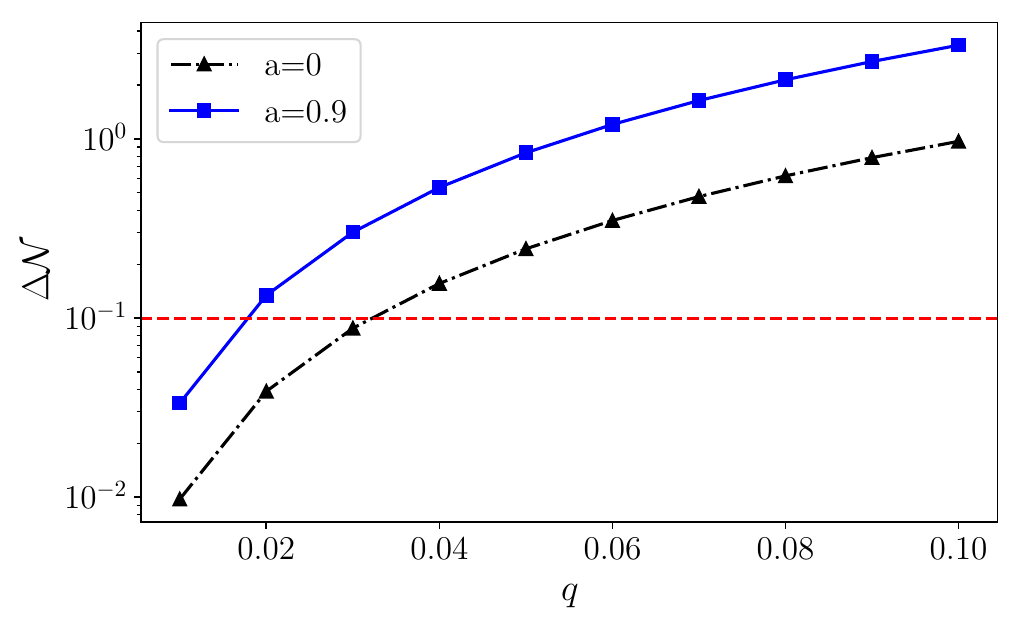}
    \caption{
    The dephasing difference $\Delta{\cal N}={\cal N}_\text{vec}-{\cal N}_\text{scal}$ caused by the vector and scalar fields for different spin $a$ as a function of charge $q_s=q_v=q$. All binaries are observed one year before the merger.}
    \label{dephasing}
\end{figure}

Hereafter, we study the EMRI system whose primary body with mass $M=10^6~M_\odot$, dimensionless spin $a=0.9$, and $a=0.0$, while the secondary body a compact object with mass $m_p=10~M_\odot$.
The initial radial position $r=p_0$ is chosen to ensure the orbital evolution after a year on a circular equatorial orbit reaches a distance of $0.1~M$ from the ISCO.
A preliminary assessment of the detectability of the charge can be made by examining the evolution of the phase of the GW signal.
The fact that the phase difference between the scalar charge and vector charge exceeds a certain threshold indicates that LISA can distinguish the scalar field and vector field after twelve months of observation.
To quantify the impact of this difference caused by scalar and vector fields on possible GW detections by future interferometers like LISA, we compute the GW phase difference accumulated before the merger \cite{Berti:2004bd}:
\begin{equation}
{\cal N}=2\pi\int_{f_{\rm min}}^{f_{\max}}\frac{f}{\dot{f}}df\ ,
\end{equation}
where $f_{\rm max}$ is the GW frequency at the ISCO and $f_{\rm min}$ is the GW frequency one year before the ISCO  \cite{Pani:2011xj}.
Figure \ref{dephasing} shows the GW phase difference $\Delta{\cal N}={\cal N}_\text{vec}-{\cal N}_\text{scal}$ for the above EMRI system as a function of charge.
The difference is always positive since the vector field emission is larger than the scalar field emission.
The dephasing difference between the scalar field and vector field for Kerr BH is larger than the Schwarzschild BH with the same charge $q_s=q_v=q$.
For the Kerr BH with $a=0.9$ and $q\geq 0.02$, the dephasing can be larger than $0.1$ radiant, corresponding to the threshold above which the two signals are distinguished by LISA, while for the Schwarzschild BH $q$  should be larger than $0.04$.

A more quantitative analysis of LISA's ability to distinguish between scalar and vector charges is the faithfulness ${\cal F}$ between two GW signals emitted by binaries with scalar charge and with vector charge (see supplementary material \ref{appendixsignal}).
Figure~\ref{faith} shows the faithfulness which estimates how much two signals differ, weighted by the noise spectral density of LISA.
For a signal with the signal-to-noise ratio (SNR) $\rho=30$, values of ${\cal F}\lesssim 0.988$ indicate that the two waveforms are significantly different and do not provide an accurate description of one another \cite{Flanagan:1997kp, Lindblom:2008cm, Chatziioannou:2017tdw}.
Figure~\ref{faith} shows the values of ${\cal F}$ for the chosen prototype of binary configuration as a function of the charge $q$.
After one year of observation with $\rho=30$, the faithfulness is always smaller than the threshold for charge $q\gtrsim 0.02$ in the background of Kerr BH with spin $a=0.9$ and $q\gtrsim 0.04$ in the background of Schwarzschild BH.
\begin{figure}
    \centering
    \includegraphics[width=0.96\columnwidth]{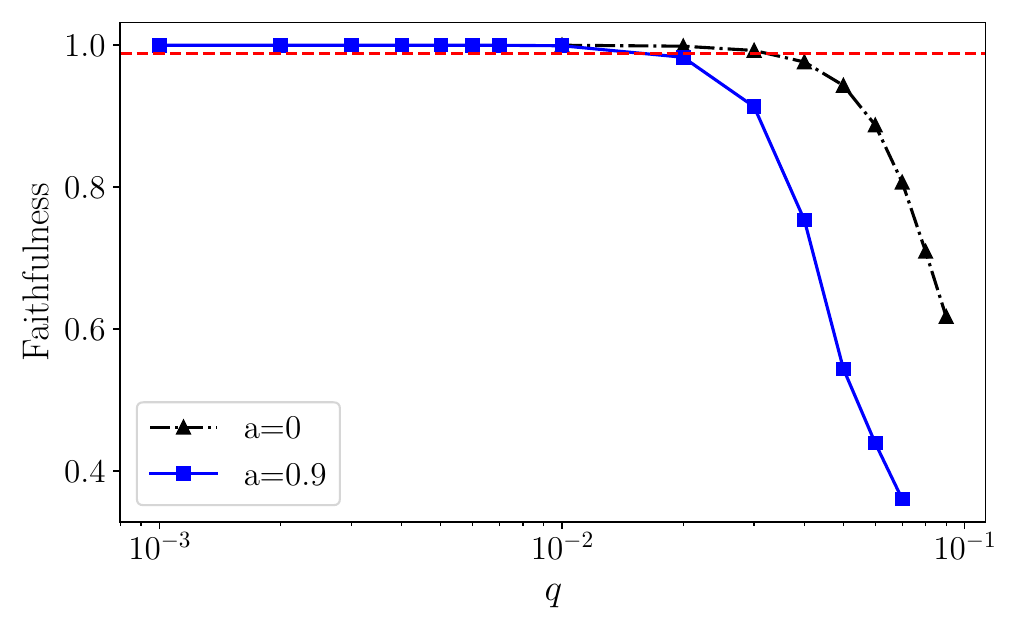}
    \caption{
    The faithfulness caused by the vector and scalar fields for different spin $a$. The horizontal dashed line represents the detection limit with LISA, $\mathcal{F}_n= 0.988$.}
    \label{faith}
\end{figure}
However, the use of dephasing and faithfulness is a rough treatment in the sense that multiple correlations are ignored. 
Such a treatment will usually result in deceptively larger effects while in reality some (or most of) effects can be absorbed into the estimation of other parameters. 
A more reliable method involves the use of parameter estimation methods, like Fisher information matrix (FIM) approximation or Bayesian analysis.

\section{Fisher information matrix and Bayesian analysis}\label{FIMB}
In the time domain, the GW signal is mainly determined by parameters
\begin{equation}
{\bm \xi}=(\ln M, \ln m_p, a, p_0,q_s, q_v, \theta_s, \phi_s, \theta_1, \phi_1, d_L),
\end{equation}
where $m_p$ and $p_0$ are the mass and radius of the smaller compact object at $t=0$.
Bayesian inference method is based on Bayes rule
\begin{equation}\label{EqBayes}
    p({\bm \xi}|d) = \frac{p(d|{\bm \xi})p({\bm \xi})}{p(d)},
\end{equation}
where $p({\bm \xi}|d)$ is the posterior distribution of the parameters ${\bm \xi}$, $p(d|{\bm \xi})$ is the likelihood,
\begin{equation}\label{EqLikelihood}
    p(d|{\bm \xi})=\exp\left[-\frac{1}{2}(h({\bm \xi})-d|h({\bm \xi})-d)\right],
\end{equation}
$d=h({\bm \xi_0})+n$ is the observed data for the true parameters ${\bm \xi_0}$,
$n$ is the noise generated by the noise power spectra,
$p({\bm \xi})$ is the prior distribution of the parameters ${\bm \xi}$,
and $p(d)$ is the evidence which is treated as a normalization constant.
In the large SNR limit,
the covariances of source parameters $\xi$  are given by the inverse of the Fisher information matrix
\begin{equation}
\Gamma_{i j}=\left\langle\left.\frac{\partial h}{\partial \xi_{i}}\right| \frac{\partial h}{\partial \xi_{j}}\right\rangle_{\xi=\hat{\xi}}.
\end{equation}
The statistical error on $\xi$ and the correlation coefficients between the parameters are provided by the diagonal and non-diagonal parts of ${\bf \Sigma}={\bf \Gamma}^{-1}$, i.e.
\begin{equation}
\sigma_{i}=\Sigma_{i i}^{1 / 2} \quad, \quad c_{\xi_{i} \xi_{j}}=\Sigma_{i j} /\left(\sigma_{\xi_{i}} \sigma_{\xi_{j}}\right).
\end{equation}
Because of the triangle configuration of the space-based GW detector regarded as a network of two L-shaped detectors, with the second interferometer rotated of $60^\circ$ with respect to the first one, the total SNR can be written as the sum of SNRs of two L-shaped detectors
\begin{equation}
\rho=\sqrt{\rho_1^2+\rho_2^2}=\sqrt{\left\langle h_1|h_1 \right\rangle+\left\langle h_2|h_2 \right\rangle},
\end{equation}
where $h_1$ and $h_2$ denote the signals detected by two L-shaped detectors.
The total covariance matrix of the source parameters is obtained by inverting the sum of the Fisher matrices $\sigma_{\xi_i}^2=(\Gamma_1+\Gamma_2)^{-1}_{ii}$.
The initial orbital separation $r_0$ is adjusted to experience one-year adiabatic evolution before the final plunge $r_{\text{end}}=r_{\text{ISCO}}+0.1~M$.
We perform the Bayesian analysis and FIM estimation for the EMRI system I with $m_p=10~M_{\odot}$, $M=10^6~M_{\odot}$, $a=0$, scalar charge $q_s=0.2$, vector charge $q_v=0.2$, and the luminosity distance $d_L$ can be changed freely to vary the SNR of the signal.
We also perform only FIM estimation for the EMRI system II with $m_p=10~M_{\odot}$, $M=10^6~M_{\odot}$, $a=0.9$, scalar charge $q_s=0.2$, vector charge $q_v=0.2$, and the luminosity distance $d_L$ can be changed freely to vary the SNR of the signal.

\begin{figure}
\centering
\includegraphics[width=0.9\columnwidth]{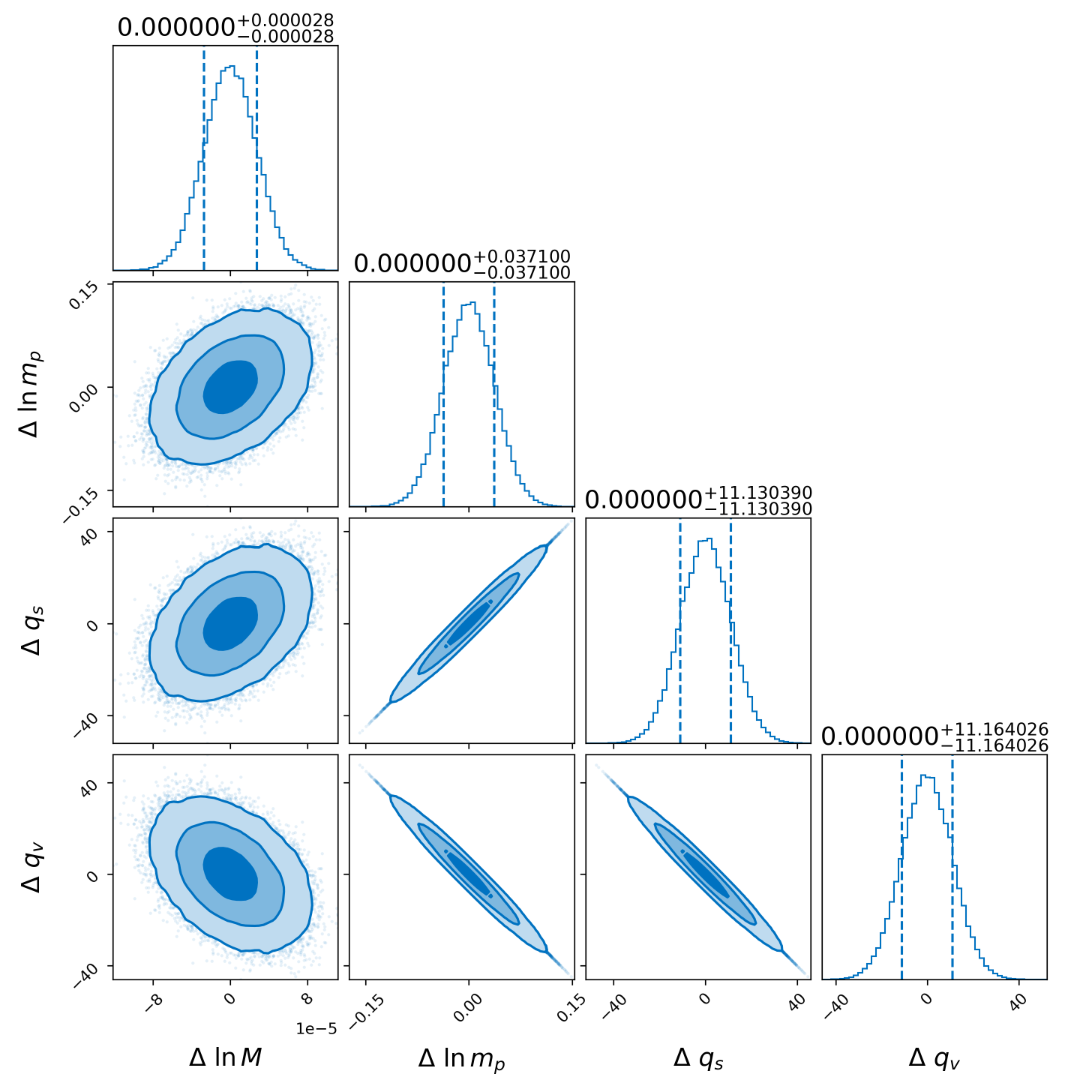}
\caption{Corner plot for the probability distribution of the source parameters $(\ln M,\ln m_p, q_s, q_v)$ with LISA, inferred after one-year observations of EMRIs with $q_s=q_v=0.2$ and $a=0$.
Vertical lines show the $1\sigma$ interval for the source parameter.
The contours correspond to the $68\%$, $95\%$, and $99\%$ probability confidence intervals.}
\label{corner00}
\end{figure}

\begin{figure}
\centering
\includegraphics[width=0.9\columnwidth]{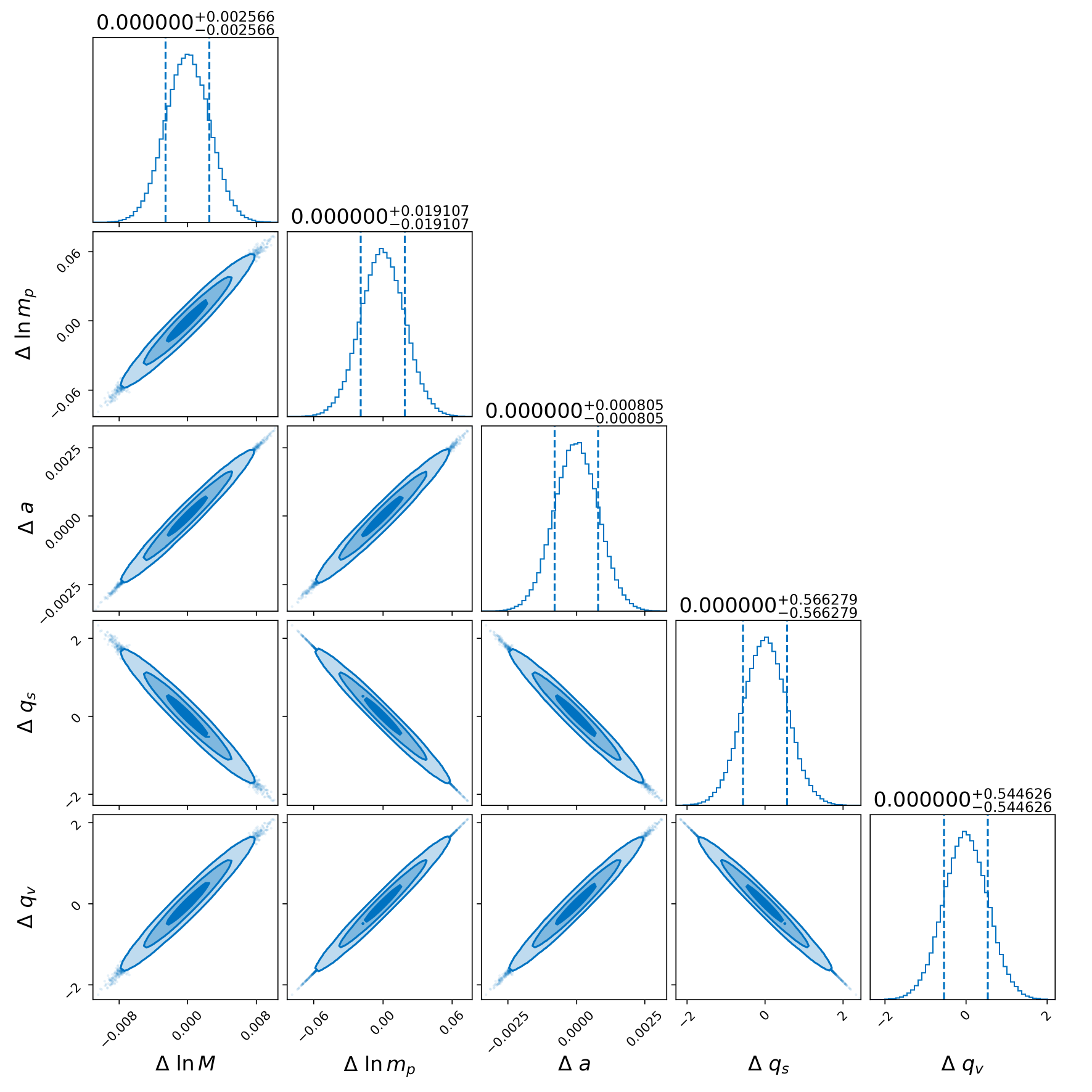}
\caption{Corner plot for the probability distribution of the source parameters $(\ln M,\ln m_p, a, q_s, q_v)$ with LISA, inferred after one-year observations of EMRIs with $q_s=q_v=0.2$ and $a=0.9$.
Vertical lines show the $1\sigma$ interval for the source parameter.
The contours correspond to the $68\%$, $95\%$, and $99\%$ probability confidence intervals.}
\label{corner09}
\end{figure}

Figures \ref{corner00} and \ref{corner09} show the probability distribution obtained by the Fisher matrix information for the binary masses, the spin of the primary, the scalar charge, and the vector charge, for EMRIs observed one year before the plunge with $q_s=q_v=0.2$ and SNR of $150$.
This analysis shows that the measurement of the scalar charge and vector charge can not be detected with a relative error smaller than $100\%$, with a probability distribution that does not have any support for distinguishing the scalar or vector charge of the secondary.
Due to the correlations that exist between the scalar and vector charge, the scalar and vector charge are not distinguishable even if they produce a nonnegligible dephasing and faithfulness.
From Figs. \ref{corner00} and \ref{corner09}, we find that the error of scalar charge and vector charge $\sigma_{q_s}\approx\sigma_{q_v}\approx11$ for EMRI I with $a=0$, while the error $\sigma_{q_s}\approx\sigma_{q_v}\approx0.5$ for EMRI II with $a=0.9$.
This result is consistent with our previous analysis that the Kerr BH background causes more difference in flux between the scalar field and vector field, thus resulting in smaller error for scalar and vector charge.
We also perform the Fisher matrix information for EMRIs observed four years before the plunge in Appendix \ref{FIMstability}.
Increasing the observation time up to four years can decrease the error of the scalar charge and vector charge.
However, the improvement of the error of the scalar charge and vector charge is not obvious.
We can only distinguish the scalar and vector charge when the charge is larger than $0.2$ and SNR of 150.

\begin{figure}
  \centering
  \includegraphics[width=0.9\columnwidth]{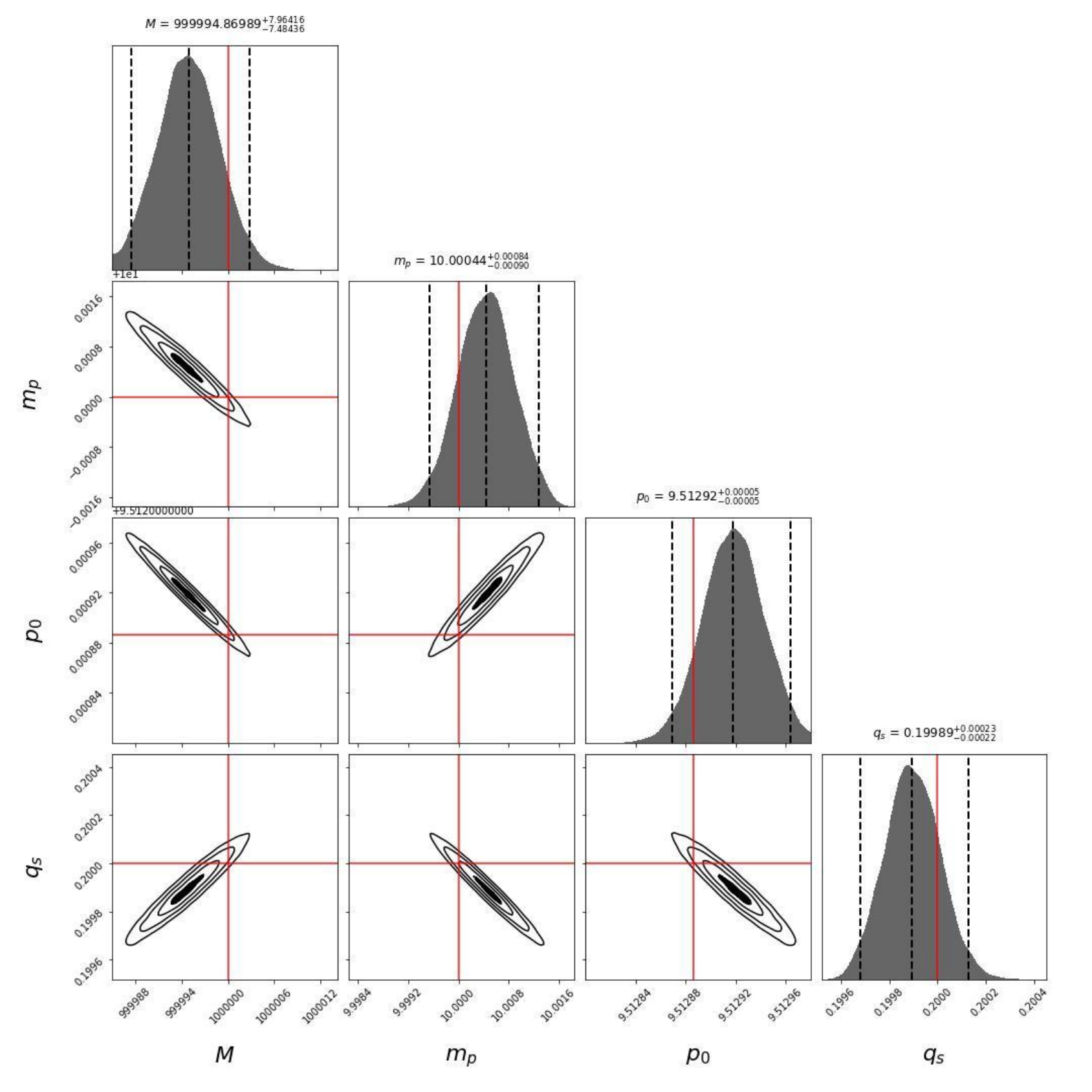}
  \caption{The parameter estimation errors using the Bayesian analysis for the simulated source using the GW model including only scalar field.}
  \label{BayesianScalar}
\end{figure}

\begin{figure}
  \centering
  \includegraphics[width=0.9\columnwidth]{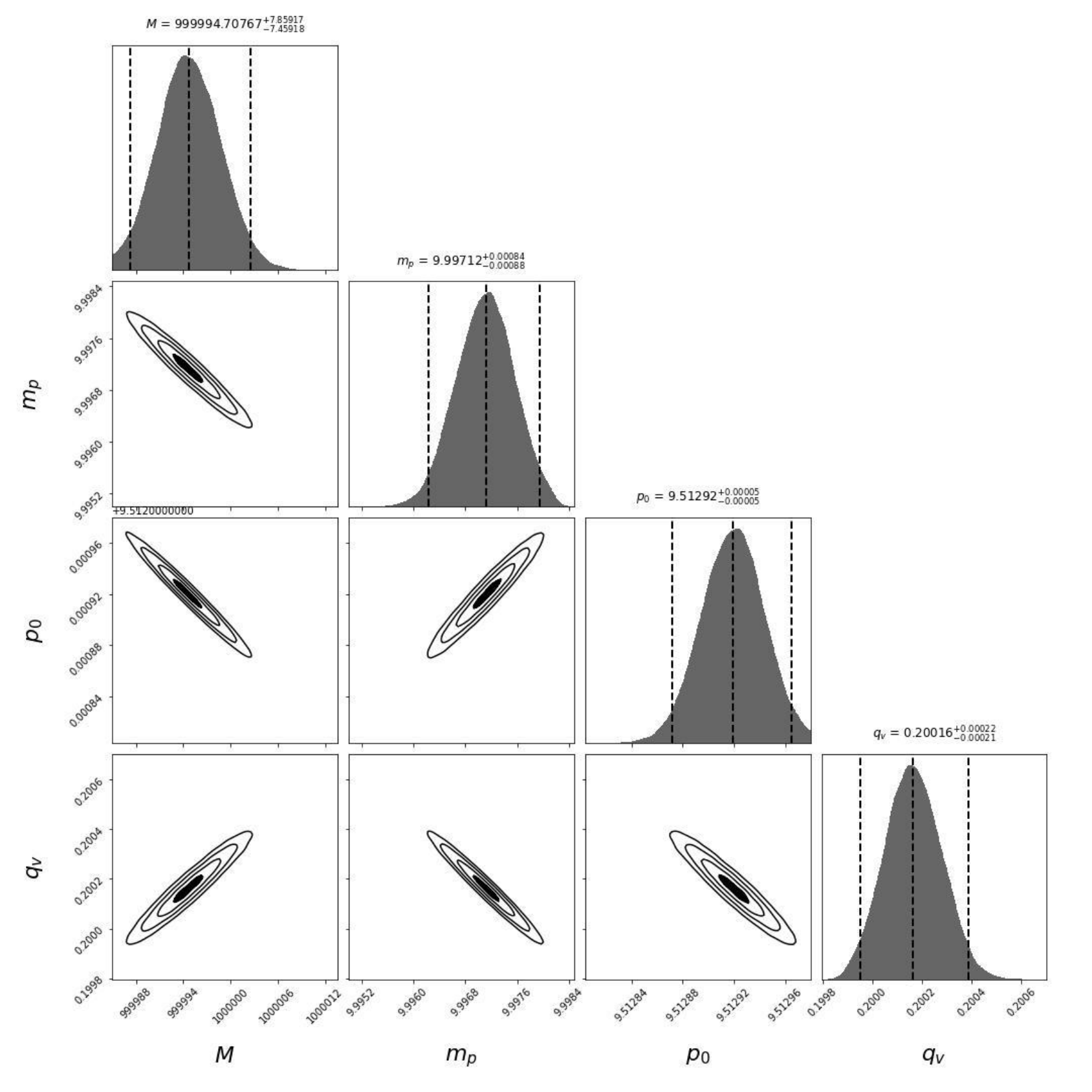}
  \caption{The parameter estimation errors using the Bayesian analysis for the simulated source using the GW model including only vector field.}
  \label{BayesianVector}
\end{figure}

To perform the Bayesian analysis, we need to get the fast waveform.
Making use of the FastEMRIWaveforms, based on the open code \cite{Katz:2021yft}, we can add the influence of the extra field into the code by adding scalar and vector flux values and get fast trajectory evolution and waveforms.
In our analysis, we simulate signals in LISA for the source with parameters: $m_p=10~M_{\odot}$, $M=10^6~M_{\odot}, a=0$, scalar charge $q_s=0.2$, distance $d_L=0.1~\rm{Gpc}$, then use the public code Bilby \cite{Ashton:2018jfp}
to perform Bayesian analyses with Eqs. \eqref{EqBayes} and ~\eqref{EqLikelihood}.
We choose the sampler Dynesty \cite{dynesty,2004AIPC..735..395S,10.1214/06-BA127}
and $1000$ live points for nested sampling and we obtain the posteriors of the physical parameter $\bm{\xi}$. 
The signal-to-noise ratio for the source is SNR=427.
The results of the parameter estimations are shown in Figs. \ref{BayesianScalar} and \ref{BayesianVector}.
We fit the scalar model including only the scalar field to the simulated data in Fig. \ref{BayesianScalar}.
We fit the vector model including only the vector field to the simulated data in Fig. \ref{BayesianVector}.
The Bayesian convergence tests can be seen in Appendix \ref{BCtest}.
To quantify whether we can distinguish the scalar field from the vector field, we take a Bayesian approach by computing the Bayes factor between the scalar model and the vector model.
A signal for which the Bayes factor (BF) exceeds $100$ can be understood as decisively favored \cite{Kass:1995loi}.
The Bayesian factor between the models with scalar and vector charge is $\text{BF}\approx 1$.
The GR systematic bias due to non-zero scalar charge has been investigated in literature \cite{Speri:2024qak}.
GR theory and modified gravity theory with scalar field can be distinguished as long as the scalar charge is larger than $0.015$.
So EMRIs can help us test fundamental fields like the scalar or vector field but are not suitable for us to distinguish scalar fields from vector fields for their correlations that exist between the scalar and vector flux.

\section{Extra polarization test}\label{EPT}
\begin{figure}
\centering
\includegraphics[width=0.9\columnwidth]{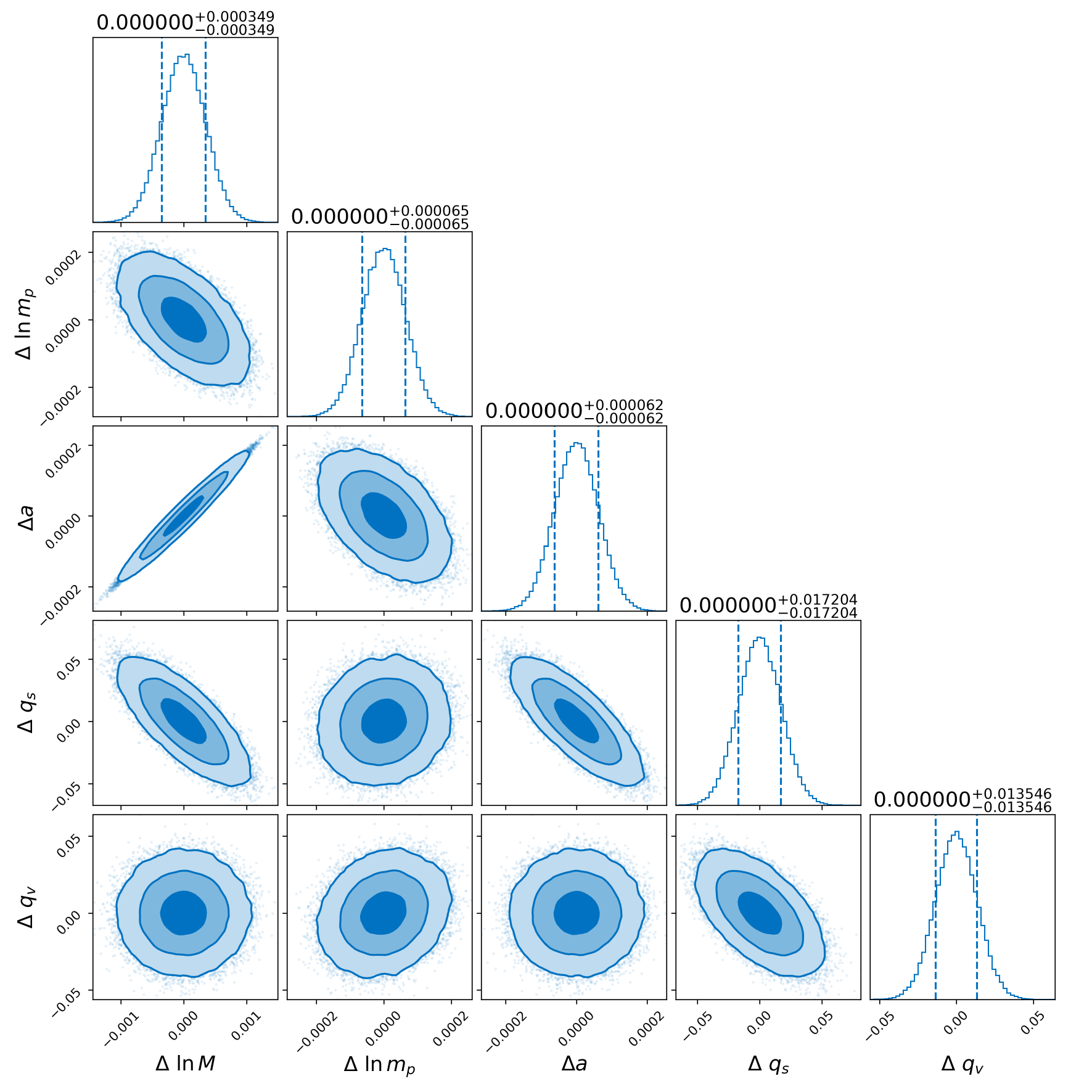}
\caption{Corner plot for the probability distribution of the source parameters $(\ln M,\ln m_p, a, q_s, q_v)$ with LISA including the effect of extra polarizations, inferred after one-year observations of EMRIs with $q_s=q_v=0.2$ and $a=0.9$.
Vertical lines show the $1\sigma$ interval for the source parameter.
The contours correspond to the $68\%$, $95\%$, and $99\%$ probability confidence intervals.}
\label{EMcorner09}
\end{figure}

Apart from the influence of flux on the orbital evolution and GW phase, scalar and vector fields can also emit extra GW polarization.
Given the source orientation $\hat{N}(\theta_s,\phi_s)$ and the orbital angular direction $\hat{L}(\theta_1,\phi_1)$, the set of unit vectors $\{\hat{m},\hat{n},\hat{\omega}\}$ can be defined 
\begin{equation}
\begin{split}
\hat{m}&=\hat{N}\times \hat{L}, \\
\hat{n}&=\frac{\hat{m}\times \hat{N}}{|\hat{m}\times \hat{N}|},\\
\hat{w}&=-\hat{N}.
\end{split}
\end{equation}
The tensor and extra polarization tensors are defined as follows
\begin{equation}
\begin{split}
e_{ij}^+&=\hat{m}_{i}\hat{m}_{j}-\hat{n}_{i}\hat{n}_{j},\qquad
e_{ij}^\times=\hat{m}_{i}\hat{n}_{j}+\hat{n}_{i}\hat{m}_{j},\\
e_{ij}^x&=\hat{m}_{i}\hat{w}_{j}+\hat{m}_{j}\hat{w}_{i},\qquad  e_{ij}^y=\hat{n}_{i}\hat{w}_{j}+\hat{n}_{j}\hat{w}_{i}\\
e_{ij}^b&=\hat{m}_{i}\hat{m}_{j}+\hat{n}_{j}\hat{n}_{i},\qquad e_{ij}^l=\hat{w}_{i}\hat{w}_{j}.
\end{split}
\end{equation}
For the massless scalar field, the amplitude of the breathing mode is \cite{Chatziioannou:2012rf,Hansen:2014ewa,OBeirne:2019lwp,Liu:2020mab}
\begin{equation}
h_b=\sqrt{2}q_s m_p/d_{L} [M \omega(t)]^{1/3} \sin(\iota ) \cos [\varphi_{\rm orb}(t)+\varphi_0].
\end{equation}
For the massless vector field, the amplitude of the vector mode is \cite{Chatziioannou:2012rf,Hansen:2014ewa,OBeirne:2019lwp,Liu:2020mab}
\begin{equation}
\begin{split}
h_x=&q_v m_p/d_{L} [M \omega(t)]^{1/3} \cos(\iota ) \cos [\varphi_{\rm orb}(t)+\varphi_0],\\
h_y=&q_v m_p/d_{L} [M \omega(t)]^{1/3}  \sin [\varphi_{\rm orb}(t)+\varphi_0].
\end{split}
\end{equation}
The GW emitted by the EMRI including scalar field and vector field is 
\begin{equation}
h_{ij}=h_+ e_{ij}^++h_\times e_{ij}^\times +h_x e_{ij}^x+h_y e_{ij}^y+h_b e_{ij}^b.
\end{equation}
The GW strain measured by the detector is
\begin{equation}
h(t)=D^{ij}h_{ij}=h_{+}(t) F^{+}(t)+h_{\times}(t) F^{\times}(t)+\sum_{I=x,y,b}h_{I}(t) F^{I}(t),
\end{equation}
and under the long-wavelength approximation the detector tensor $D^{ij}$ is 
\begin{equation}
D^{ij}=\frac{1}{2}\left[\hat{u}^i \hat{u}^j-\hat{v}^i\hat{v}^j\right],
\end{equation}
where $\hat{u}$ and $\hat{v}$ are the unit vectors along the arms of the detector.
The response function for extra polarizations $F^{I=x, y, b}=D^{ij}e^I_{ij}$ can be seen in Ref. \cite{Liang:2019pry}.
Including the effect of extra polarizations, we perform the FIM estimation for the EMRI system II with $m_p=10~M_{\odot}$, $M=10^6~M_{\odot}$, $a=0.9$, scalar charge $q_s=0.2$, vector charge $q_v=0.2$, and the luminosity distance $d_L$ can be changed freely to vary the SNR of the signal.
Figure \ref{EMcorner09} shows the probability distribution obtained by the Fisher matrix information for the binary masses, the spin of the primary, the scalar charge, and the vector charge, for EMRIs observed one year before the plunge with $q_s=q_v=0.2$ and SNR of $150$.
We can see from Fig. \ref{EMcorner09} that the error for scalar charge is $\sigma_{q_s}=0.017$ and for vector charge is $\sigma_{q_v}=0.014$.
This result still holds even for the charge $q_s=q_v=0$.
So the error limit for scalar and vector charge is approximately $0.01$ for the EMRI system II.
Including extra polarizations can break the correlations between the scalar field and vector field.

\section{Conclusion}\label{conclusion}
In summary, EMRIs are golden binaries to test fundamental fields.
Modified gravity theories, including scalar or vector fields, can not be distinguished through GW observations in the weak-field regime since the energy fluxes of scalar and vector waves share the same behavior up to $O(v^3)$.
The vector and scalar flux behavior is different in a strong-field regime whose ratio can exceed $1$ at most $2\%$.
Based on the analysis of Bayesian analysis and Fisher information matrix, our results indicate that the scalar field and vector field surrounded by the smaller BH carrying charge are not suitable to be distinguished by the EMRI system for LISA if we only consider the influence of scalar and vector flux on the orbital evolution and tensor GW phase.
However, including extra polarizations can break the correlations between the scalar field and vector field and then help us distinguish the scalar and vector field if the charge is approximately above $0.01$.

\begin{acknowledgments}
This work was supported in part by the National Key Research and Development Program of China under Grant No. 2020YFC2201504, and the China Postdoctoral Science Foundation under Grant No. 2023M742297.
\end{acknowledgments}

\appendix
\section{Signal for LISA}\label{appendixsignal}
Having computed the emitted energy flux, the EMRI's adiabatic evolution is determined by
 \begin{equation}\label{orbittime}
 \frac{d r}{dt}=-\dot{E}\left(\frac{d E_{\text{orb}}}{dr}\right)^{-1},\qquad \frac{d \varphi_{\text{orb}}}{d t}=\pi f,
 \end{equation}
 where 
\begin{equation}
E_{\text{orb}}= m_p\frac{r^{3 / 2}-2 r^{1 / 2} + a }{r^{3 / 4}\left(r^{3 / 2}-3  r^{1 / 2} + 2 a \right)^{1 / 2}}    
\end{equation}
 and
$\dot{E}=\dot{E}_{\text{grav}}+\dot{E}_{\text{vec}}+\dot{E}_{\text{scal}}$.
We can obtain the inspiral trajectory from adiabatic evolution in Eq. \eqref{orbittime}, then compute GWs in the quadrupole approximation.
The metric perturbation in the transverse-traceless (TT) gauge is
\begin{equation}
h_{i j}^{\mathrm{TT}}=\frac{2}{d_L}\left(P_{i l} P_{j m}-\frac{1}{2} P_{i j} P_{l m}\right) \ddot{I}_{l m},
\end{equation}
where $d_L$ is the luminosity distance of the source,
$P_{ij}=\delta_{ij}-n_i n_j$ is the projection operator acting onto GWs with the unit propagating direction $n_j$,
$\delta_{ij}$ is the Kronecker delta function,
and $\ddot{I}_{ij}$ is the second time derivative of the mass quadrupole moment.
The GW strain measured by the detector is
\begin{equation}\label{signal}
h(t)=h_{+}(t) F^{+}(t)+h_{\times}(t) F^{\times}(t),
\end{equation}
where $h_+(t)=\mathcal{A}\cos\left[2\varphi_{\rm orb}+2\varphi_0\right]\left(1+\cos^2\iota\right)$, $h_\times(t)=-2\mathcal{A}\sin\left[2\varphi_{\rm orb}+2\varphi_0\right]\cos\iota$,  $\iota$ is the inclination angle between the binary orbital angular momentum and the line of sight,
the GW amplitude  $\mathcal{A}=2m_{\rm p}\left[M\omega(t)\right]^{2/3}/d_L$ and $\varphi_0$ is the initial phase.
The interferometer pattern functions $F^{+,\times}(t)$ and $\iota$ can be expressed in terms of four angles which specify the source orientation, $(\theta_s,\phi_s)$,
and the orbital angular direction $(\theta_1,\phi_1)$.
The source angles are set as $\theta_s=\pi/5,~\phi_s=\pi/6$ and $\theta_1=\pi/3,~\phi_1=\pi/4$.
The faithfulness between two signals is defined as
\begin{equation}\label{eq:def_F}
\mathcal{F}[h_1,h_2]=\max_{\{t_c,\phi_c\}}\frac{\langle h_1\vert
	h_2\rangle}{\sqrt{\langle h_1\vert h_1\rangle\langle h_2\vert h_2\rangle}}\ ,
\end{equation}
where $(t_c,\phi_c)$ are time and phase offsets \cite{Lindblom:2008cm},
the noise-weighted inner product between two templates $h_1$ and $h_2$ is
\begin{equation}\label{product}
\left\langle h_{1} \mid h_{2}\right\rangle=4 \Re \int_{f_{\min }}^{f_{\max }} \frac{\tilde{h}_{1}(f) \tilde{h}_{2}^{*}(f)}{S_{n}(f)} df,
\end{equation}
$\tilde{h}_{1}(f)$ is the Fourier transform of the time-domain signal $h(t)$,
its complex conjugate is $\tilde{h}_{1}^{*}(f)$,
and $S_n(f)$ is the noise spectral density for space-based GW detectors.
The explicit formula of the noise spectral density for LISA composed of the instrumental and the confusion noise produced by
unresolved galactic binaries are given in \cite{Maselli:2021men,Robson:2018ifk}.
The signal-to-noise ratio (SNR) can be obtained by calculating $\rho=\left\langle h|h \right\rangle^{1/2}$.
As pointed out in \cite{Chatziioannou:2017tdw},
two signals can be distinguished by LISA if $\mathcal{F}_n\leq0.988$.
In order to get the fast waveform, we make use of the improved AAK waveform of FastEMRIWaveforms.
For the equatorial circular orbit, the GW strain measured by LISA in FastEMRIWaveforms is identical to the formula given by Eq. \eqref{signal}.

\section{Bayesian convergence}\label{BCtest}
\begin{figure}
\centering
\includegraphics[width=0.9\columnwidth]{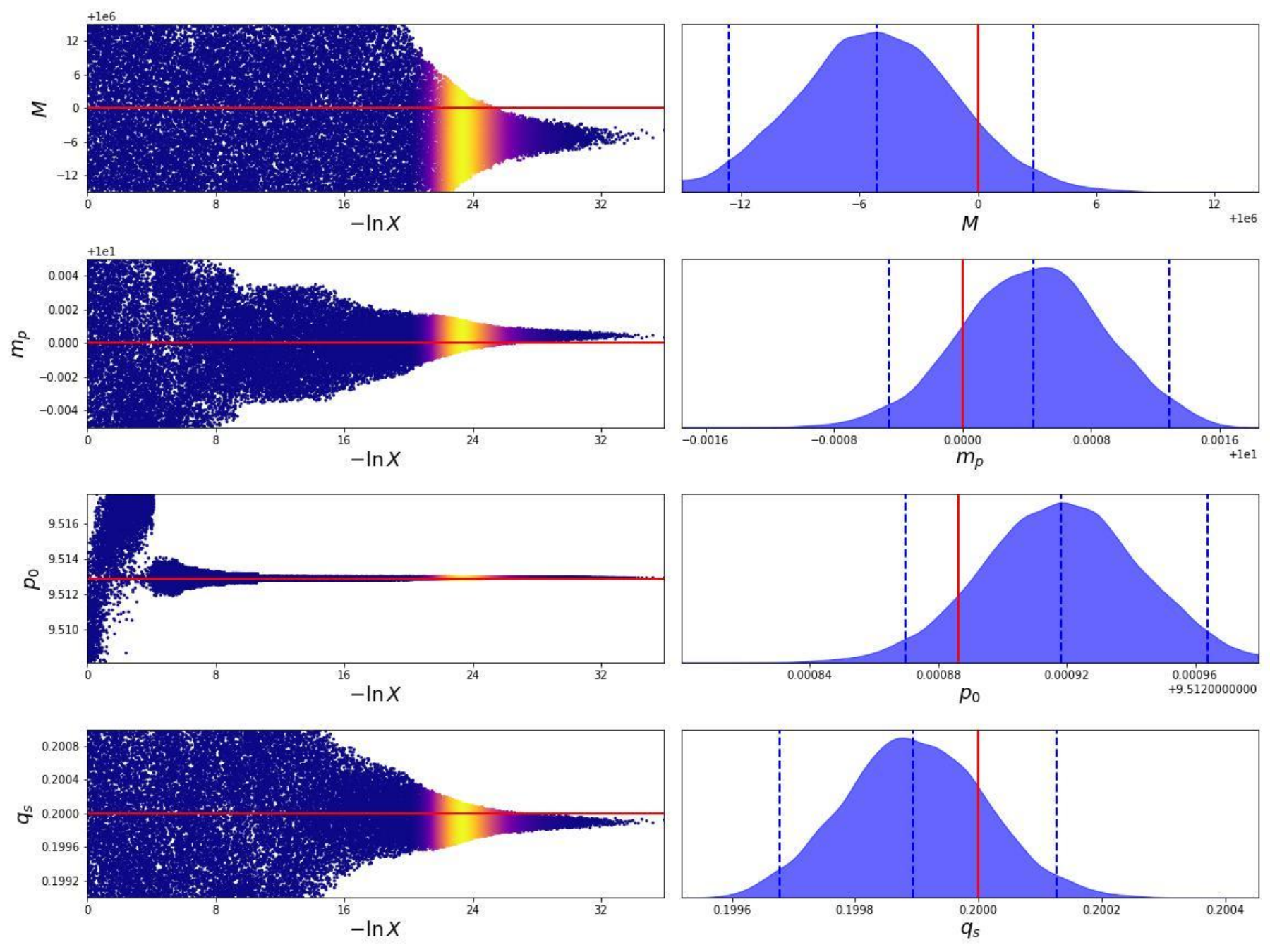}
\caption{Trace plot for the parameter estimation errors using the Bayesian analysis for the simulated source using the GW model including only scalar field.}
\label{tracescalar}
\end{figure}

\begin{figure}
\centering
\includegraphics[width=0.9\columnwidth]{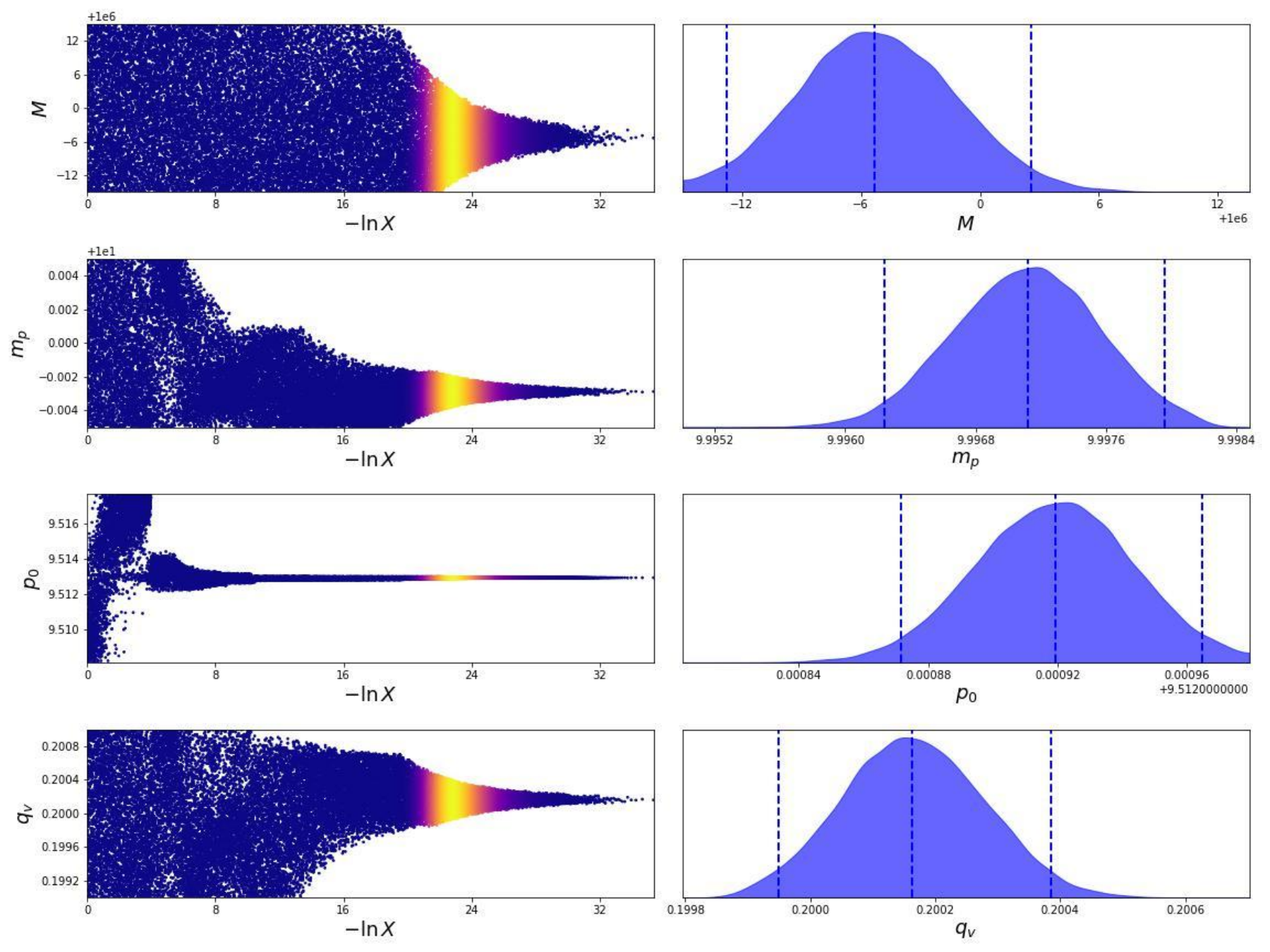}
\caption{Trace plot for the parameter estimation errors using the Bayesian analysis for the simulated source using the GW model including only vector field.}
\label{tracevector}
\end{figure}

To test whether the Bayesian sampling has fully converged, we give the trace plot of the Bayesian results.
Figure \ref{tracescalar} shows the positions of samples over the course of the run for the Bayesian result of Fig. \ref{BayesianScalar}.
Figure \ref{tracevector} shows the positions of samples over the course of the run for the Bayesian result of Fig. \ref{BayesianVector}.
All intrinsic parameters have fully converged.

\section{Fisher information matrix stability}\label{FIMstability}
In this section, we shall provide technical details on the method of the Fisher information matrix in our previous literature \cite{Zhang:2024ugv}.
The signal can be written in analytic form 
\begin{equation}
h(t)=h(t,r(t),\varphi_{\rm orb}(t),\xi)
\end{equation}
where
\begin{equation}
{\xi}=(\ln M, \ln m_p, a, p_0,q_s, q_v, \theta_s, \phi_s, \theta_1, \phi_1, d_L).
\end{equation}
We computed derivatives for the Fisher matrix with analytic expression.
Taking mass parameter $M$ for example, we can get 
\begin{equation}\label{FIMC3}
\partial_{\ln M} h(t)=\frac{\partial h}{\partial r(t)} \frac{\partial r(t)}{\partial \ln M}+\frac{\partial h}{\partial \varphi_{\rm orb}(t)} \frac{\partial \varphi_{\rm orb}(t)}{\partial \ln M}+\frac{\partial h}{\partial \ln M}.
\end{equation}
So the accuracy of the Fisher matrix is determined by the orbital evolution derivatives with respect to the parameter $M$ because the other steps are all analytical.
We can write the orbital evolution equation in a compact form
\begin{equation}
 \dot{\vec y}=\vec F (\vec y(t,\xi),t,\xi), 
\end{equation}
where $\vec y=(r(t,\xi),\varphi_{\rm orb}(t,\xi))$.
We can rewrite the formal orbital evolution equation in the form
\begin{equation}\label{eq2}
\frac{\partial \vec y(t,\xi)}{\partial t}  =  \vec F (\vec y (t,\xi),t,\xi).
\end{equation}
We can differentiate Eq. \eqref{eq2} with respect to $M$
\begin{equation}
\frac{\partial}{\partial M}    \frac{\partial \vec y(t,\xi)}{\partial t} =\frac{\partial}{\partial M} \vec F (\vec y (t,\xi),t,\xi).
\end{equation}
Using the chain rule, we can get 
\begin{equation}\label{eq3}
\frac{\partial}{\partial t} \frac{\partial \vec y(t,\xi)}{\partial M}=\frac{\partial \vec F (\vec y (t,\xi),t,\xi)}{\partial y_i}  \frac{\partial y_i}{\partial M}+ \frac{\partial \vec F(\vec y, t,\xi)}{\partial M}
\end{equation} 
The orbital evolution derivatives with respect to the parameter $M$ can be solved from 
 Eq.~\eqref{eq3}.
 Then the signal differentiation to $M$ can then be calculated by using the analytical form in Eq. \eqref{FIMC3}.
 Finally, we can get the signal $h(t)$ differentiation to $M$ in the time domain sampled with the step $\Delta t=20$ seconds and the Fisher matrix.

\begin{figure}
\centering
\includegraphics[width=0.9\columnwidth]{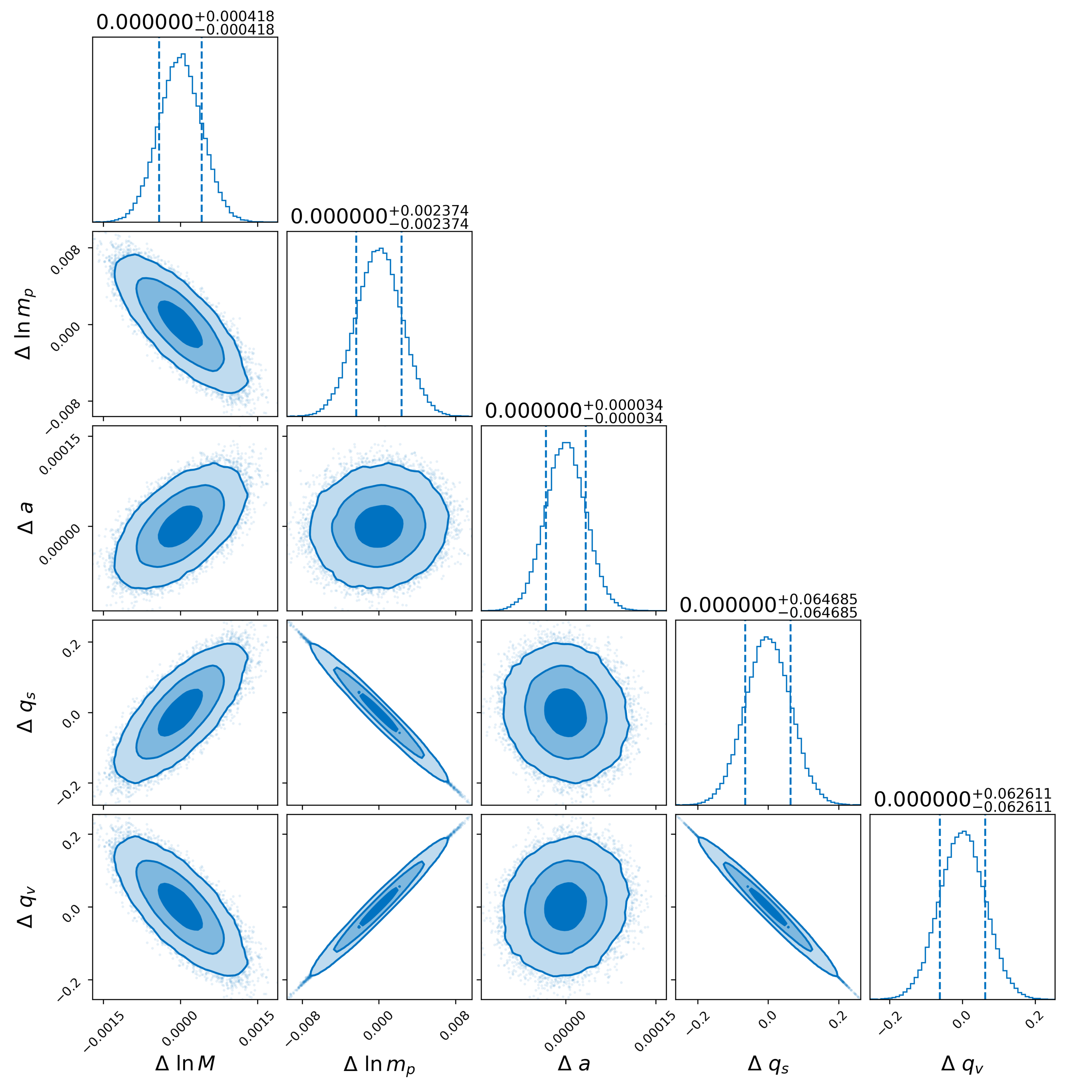}
\caption{Corner plot for the probability distribution of the source parameters $(\ln M,\ln m_p, a, q_s, q_v)$ with LISA, inferred after four-year observations of EMRIs with $q_s=q_v=0.2$ and $a=0.9$.
Vertical lines show the $1\sigma$ interval for the source parameter.
The contours correspond to the $68\%$, $95\%$, and $99\%$ probability confidence intervals.}
\label{corner09d024y}
\end{figure}

\begin{figure}
\centering
\includegraphics[width=0.9\columnwidth]{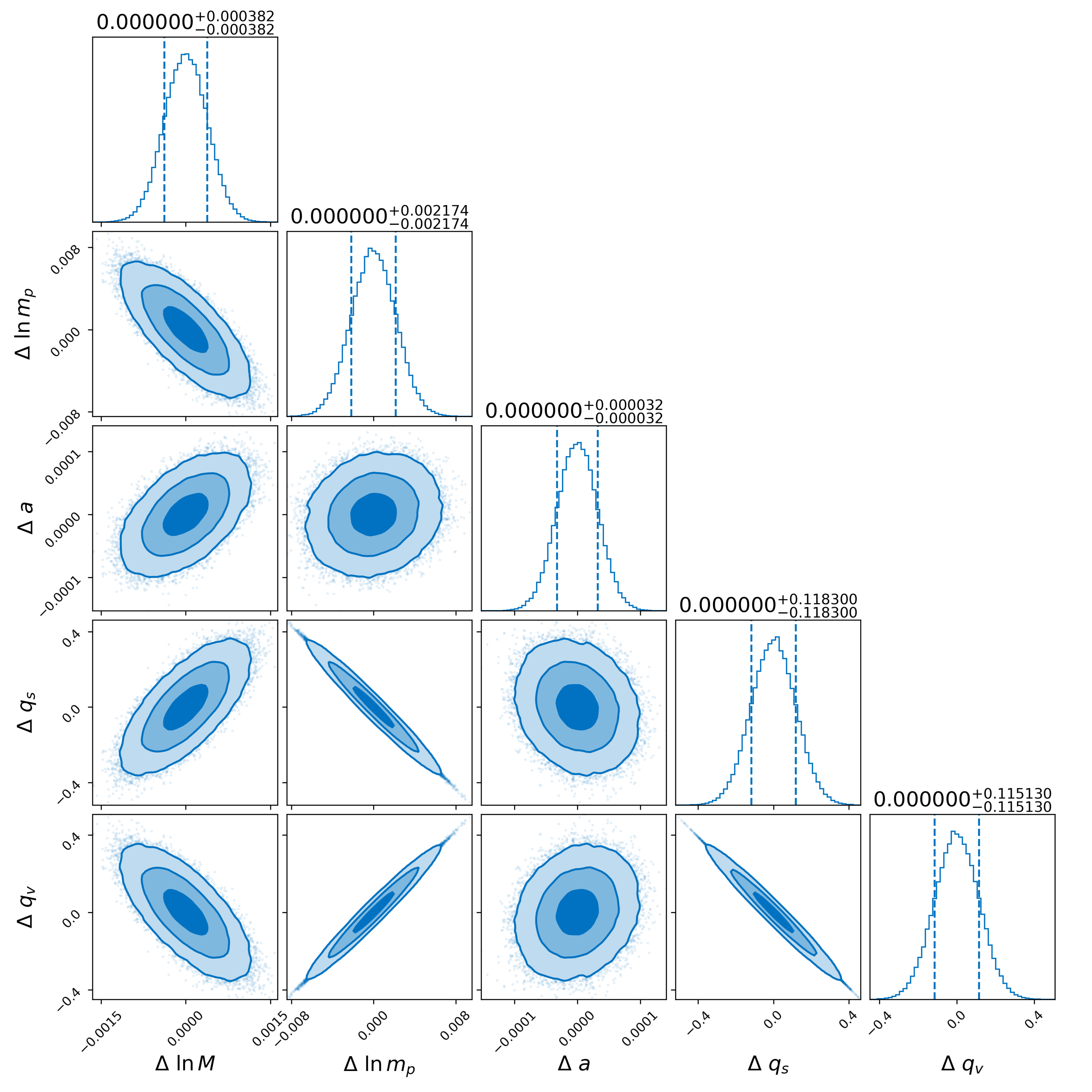}
\caption{Corner plot for the probability distribution of the source parameters $(\ln M,\ln m_p, a, q_s, q_v)$ with LISA, inferred after four-year observations of EMRIs with $q_s=q_v=0.1$ and $a=0.9$.
Vertical lines show the $1\sigma$ interval for the source parameter.
The contours correspond to the $68\%$, $95\%$, and $99\%$ probability confidence intervals.}
\label{corner09d014y}
\end{figure}

We use the method in Refs. \cite{Piovano:2021iwv,Zi:2023pvl,Zi:2024jla} to assess the stability of the FIM.
The stability can be quantitatively given by
\begin{equation}
\delta_{\rm stability}=\max_{ij}\left[\frac{((\Gamma+F)^{-1}-\Gamma^{-1})_{ij}}{\Gamma^{-1}_{ij}}\right],
\end{equation}
where $F_{ij}$ is the deviation matrix whose elements are uniform distribution around $[u_0,u_1]$.
The results of stability $\delta_{\rm stability}$ can be seen in Table \ref{stability}.
\begin{table}
  \centering
  	\begin{tabular}{|c|c|c|c|c|}
		\hline
  \multicolumn{5}{|c|}{One-year observation} \\ \hline
   \multicolumn{3}{|c|}{Without extra polarizations}&\multicolumn{2}{|c|}{Including extra polarizations} \\ \hline
$[u_0,u_1]$ & $a=0$ & $a=0.9$ &$[u_0,u_1]$ & $a=0.9$\\ \hline
$[-10^{-7},10^{-7}]$ & $5.8\times 10^{-6}$ & $2.7\times 10^{-7}$ &$[-10^{-7},10^{-7}]$ &$6.3\times 10^{-8}$\\ \hline
$[-10^{-5},10^{-5}]$& $7.1\times 10^{-4}$ & $1.4\times 10^{-4}$ &$[-10^{-5},10^{-5}]$ & $5.2\times 10^{-6}$\\ \hline
\multicolumn{5}{|c|}{Four-year observation} \\ \hline
\multicolumn{3}{|c|}{Without extra polarizations}&\multicolumn{2}{|c|}{Including extra polarizations} \\ \hline
$[u_0,u_1]$ & \multicolumn{2}{|c|}{$a=0.9$} &$[u_0,u_1]$ & $a=0.9$\\ \hline
$[-10^{-7},10^{-7}]$ & \multicolumn{2}{|c|}{$1.5\times 10^{-6}$} &$[-10^{-7},10^{-7}]$&$2.3\times 10^{-7}$\\ \hline
$[-10^{-5},10^{-5}]$& \multicolumn{2}{|c|}{$7.1\times 10^{-5}$} &$[-10^{-5},10^{-5}]$ &$3.1\times 10^{-5}$\\ \hline
\end{tabular}
    \caption{The stability $\delta_{\rm stability}$ of FIM with different EMRI configuration.
    }
    \label{stability}
\end{table}
Figures \ref{corner09d024y} and \ref{corner09d014y} show the probability distribution obtained by the Fisher matrix information for binary masses, the spin of the primary, the scalar charge, and the vector charge, for EMRIs observed four years before the plunge with $q_s=q_v=0.2$ and $q_s=q_v=0.1$, respectively.
The luminosity distance $d_L$ can be changed freely to set SNR=150.
From Figs. \ref{corner09d024y} and \ref{corner09d014y}, we find that the error of scalar charge and vector charge $\sigma_{q_s}\approx\sigma_{q_v}=0.06$ for the case $q_s=q_v=0.2$, and $\sigma_{q_s}\approx\sigma_{q_v}=0.12$ for the case $q_s=q_v=0.1$.
So increasing the observation time up to four years can decrease the error of the scalar charge and vector charge.
However, the improvement of the error of the scalar charge and vector charge is not obvious.
We can only distinguish the scalar and vector charge when the charge is larger than $0.2$ and SNR of 150.
The constraint only from the flux difference between scalar and vector field without extra polarizations is worse than the constraint including extra polarizations. 


%

\end{document}